\documentclass[numberedappendix,onecolumn]{emulateapj} 

\usepackage{latexsym}   %Simbolos de latex
\usepackage{amsfonts}   %Fuentes de la AMS
\usepackage{amstext}    %Texto de la AMS
\usepackage{amsmath,amssymb}
\usepackage{mathtools}
\usepackage{braket}
\usepackage{bm}
\usepackage{empheq} 
\usepackage{MnSymbol} 
\usepackage{graphicx}
\slugcomment{}
\begin{document}
\title{Chromospheric diagnosis with Ca {\sc ii} lines: \\forward modeling in forward scattering (I). }

 \author{E.S. Carlin\altaffilmark{1}, A. Asensio Ramos\altaffilmark{2,3}} 

\altaffiltext{1}{Istituto Ricerche Solari Locarno, 6600, Locarno, Switzerland}  
\email{escarlin@irsol.es}
  \altaffiltext{2}{Instituto de Astrof\'{\i}sica de Canarias, 38205, La Laguna, Tenerife, Spain}
  \altaffiltext{3}{Departamento de Astrof\'\i sica, Facultad de F\'\i sica, Universidad de La Laguna, Tenerife, Spain}

\begin{abstract}
This paper shows the first synthetic tomography of the quiet solar chromosphere formed by spatial maps of scattering polarization. It has been calculated for the Ca{\sc ii} $8498$, $8542$ and $3934$ {\AA} lines by solving the NLTE (non-local thermodynamical equilibrium) RT (radiative transfer) problem of the second kind in a 3D atmosphere model obtained from realistic MHD (magneto-hydrodynamical) simulations. Maps of circular polarization were calculated neglecting atomic polarization. Our investigation focuses on the linear polarization signals induced by kinematics, radiation field anisotropy and Hanle effect in forward-scattering geometry. Thus, instead of considering slit profiles at the limb as normally done in the study of the second solar spectrum, we synthetize and analyze spatial maps of polarization at disk center. It allows us to understand the spatial signatures of dynamics and magnetic field in the linear polarization for discriminating them observationally. Our results suggest new ideas for chromospheric diagnosis that will be developed throughout a serie of papers. In particular, Hanle Polarity Inversion Lines and dynamic Hanle diagrams are two concepts introduced in the present work. We find that chromospheric dynamics and magnetic field topology create spatial fingerprints in the polarization maps that trace the dynamic situation of the plasma and the magnetic field. Based on such spatial features we reconstruct the magnetic field intensity in the middle chromosphere along grooves of null linear polarization. We finally address the problems of diagnosing Hanle saturation and kinematic amplification of scattering signals using Hanle diagrams.
\end{abstract}


% %%%%%%%%%%%%%%%%%%%%%%%%%%%%%%%%%%%%%%%%
\section{Introduction}
% %%%%%%%%%%%%%%%%%%%%%%%%%%%%%%%%%%%%%%%%
 
When the line of sight points to the solar disk center, i.e. in forward scattering geometry, the Hanle effect can create linear polarization (LP) in the presence of magnetic fields \citep{Trujillo-Bueno:2003aa}. Namely, a resolved magnetic field that is inclined with respect to the solar radial produces a (``right-left'') symmetry breaking in the scattering process and thus creates Q and U signals quantifying the magnetic state of the plasma. As in a general case, these signals strongly react to vertical variations of the radiation field anisotropy (``up-down'' symmetry breakings). Hence, they also contain unique footprints of the dynamic state of the atmosphere because the radiation anisotropy is highly sensitive to it. Indeed they are expected to be largely modulated by vertical gradients of temperature \citep{Trujillo-Bueno:2001aa, Carlin:2013aa} and velocity \citep{carlin12}. This is important in spectral line polarization forming in non-LTE conditions because in such case the atomic polarization is strongly coupled with the radiation field, which is Doppler-shifted by the velocity field. Upon these basis we seek to understand the behaviour of the linear polarization (LP) signals emerging in forward scattering geometry from the strongly-dynamic and weakly-magnetized external layers of the Sun.

Nowadays, the interest of using the forward-scattering Hanle effect for magnetic field diagnosis is recent and not yet exploited. Early reports were given by \cite{Trujillo-Bueno:2002a} and \cite{Stenflo:2003} on chromospheric observations of He {\sc i} $10830$ {\AA} and Ca {\sc i} $4227$ {\AA}, respectively. Lately, \cite{Anusha:2011aa} modelled the polarization of Ca {\sc i} $4227$ {\AA} for explaining some of the observations done by \cite{michele2011cai} at disk center. In this paper we investigate the polarization in the Ca {\sc ii} IR triplet lines and in the cores of the corresponding K line considering kinematics as an indispensable new ingredient to model it.

Complete redistribution in frequencies (CRD) is a good approximation to synthetize the non-resonant Ca {\sc ii} IR triplet lines ($8498$, $8542$, $\,8662 $ {\AA}) and also the core (specially the central six Doppler widths) of the UV $3969$ and $3934$ {\AA} lines because such wavelengths are strongly affected by Doppler redistribution in the observer reference frame \citep{Uitenbroek:1989aa,Mihalas:1978}. Moreover, when the line of sight (LOS) approaches to disk center the increasing symmetry in the scattering cancels out partial redistribution effects that dominated the linearly polarized spectral wings in the CaII H and K lines towards the limb \citep{Stenflo:2006aa}. This observational fact reinforces our CRD treatment and a line-core-based diagnosis.

 Other key point to describe the line-core polarization is the saturation of the Hanle effect, by which a sufficiently strong magnetic field makes quantum coherences to vanish in the magnetic field reference frame. This produces Stokes Q and U that are insensitive to the magnetic field strength, extending the applicability of the Hanle effect to the inference of the magnetic field orientation. For a transition with polarizable upper level, Hanle saturation occurs when the effective Larmor frequency is few times the spontaneous emission rate $\mathrm{A_{u{\ell}}}$ defining the energy level uncertainty. Thus, as the Zeeman splitting is the Larmor
frequency in wavelength units, Hanle signals appear and saturate as soon as the Zeeman splitting is few times
the \textit{natural} width of the level \citep[][Sect 5.16]{ll04}. Furthermore, as weak fields give small splittings in relation to the chromospheric \textit{thermal} line widths, the transversal-Zeeman signals are still negligible and the LP remains controlled by scattering processes \citep{Landi-DeglInnocenti:1973, Jefferies:1989aa}.
That is why the the Hanle effect can be saturated with weak fields and why scattering polarization signals are essential for studying weakly-magnetized plasmas. %The larger is the difference between both widths for a given line, the more precise will be this treatment.

 For the Ca {\sc ii} IR triplet lines and the model chromosphere considered here the magnetic field is weak enough for the Hanle effect to be dominant, but strong enough to push it into saturation. This is also the most probable situation in the quiet solar chromosphere \citep{manso10}.
 The predominance of the Hanle effect is stronger in non-resonant chromospheric lines such as Ca{\sc ii} $8542$ and $\,8662 $ {\AA} because their absorption in Q and U originates completely in middle chromosphere layers, where the magnetic field intensity is reduced in comparison with lower layers. The idea is also valid in the core of resonant lines such as the Ca {\sc ii} H and K lines because they form in the even less magnetized upper chromosphere. Although in these cases Hanle clearly dominate the quiet Sun line cores, the linear polarization is in general a non-linear superposition of Hanle and Zeeman effects along wavelength.

On the contrary, the circular polarization is fundamentally described by the longitudinal Zeeman effect (wavelength splitting due to the magnetic field component along the LOS). Hence, we followed here the usual approach of disregarding the small contribution of the atomic polarization to Stokes V in order to also calculate circular polarization maps. Stokes V in the Ca {\sc ii} $8542$ and $8662$ {\AA} lines has a specially good sensitivity for diagnosing longitudinal magnetic fields \citep[e.g.,][]{Socas-Navarro:2006}.

We emphasize that the traditional modelling of scattering polarization signals and Hanle effect has always been done neglecting macroscopic velocities and studying spatial variations just along a slit at the solar limb. Although this is enough for explaining basic features of the second solar spectrum \citep[e.g., ][]{Stenflo:1997}, it appears questionable for a true diagnosis of chromospheric signals. On one hand, the action of vertical velocity gradients is able to destroy LP in very specific circumstances as well as to boost its amplitudes until more than one order of magnitude in relation to the static case \citep{Carlin:2013aa}. This becomes crucial for measuring chromospheric magnetic fields and for explaining several signals of the second solar spectrum.  On the other hand, the approach followed so far to synthetize scattering polarization is incompatible with using state-of-the-art MHD models, where dynamics and spatial extension are essential. The technical asynchrony between both fields of research is partially due to intrinsic difficulties found in the theoretical \citep{Bommier:1997ab, Casini:2014aa} and numerical \citep{Trujillo-Bueno:1999aa, Anusha:2011ab} aspects of the scattering polarization. The numerical situation worsens when considering the abrupt velocity gradients in the rarified solar chromosphere because of the relation between opacity inhomogeneities induced by Doppler shifts along a given ray and 
%Doppler-induced inhomogeneities (gaps) in the opacity for a given ray 
the (lack of) convergence to a joint solution for the rate and transfer equations \citep{Mihalas:1978}. 

We address the above-commented issues solving the RT with polarization in a radiation 3D MHD model, considering each vertical as a plane-parallel atmosphere but using the corresponding atomic populations computed in 3D by \cite{Leenaarts:2009}.  This method is reasonable because, in the chromosphere, the atomic polarization  is dominated by enhancements of radiation field anisotropy due to vertical gradients of density, temperature and velocity and not by the inhomogeneous horizontal illumination, which is azimuthally isotropized by large photon-mean-free paths and random short-scale intensity distributions troughout the horizontal plane. This paper then represents the first detailed computation and analysis of chromospheric polarization maps produced by Hanle effect and vertical variations in all the physical quantities. Possible limitations in this approach are compensated with the reduced requeriments in time and computational resources with respect to the 3D polarized-RT codes starting to emerge right now \citep{Anusha:2011ab,Stepan:2013aa}. Thus we can offer a background/guide to test the reliability of non-full-3D RT codes with polarization at the same time than developing ideas for chromospheric diagnosis.

% %%%%%%%%%%%%%%%%%%%%%%%%%%%%%%%%%%%%%%%%
\section{Technical details about the calculations}\label{sec:mhdmodels}
% %%%%%%%%%%%%%%%%%%%%%%%%%%%%%%%%%%%%%%%%
The atmospheric model used to perform the spectral synthesis is a snapshot of a radiation MHD simulation of the solar atmosphere computed by \cite{Leenaarts:2009} with the Oslo Stagger Code \citep{Hansteen:2007}. This code solves the set of MHD equations that describe the plasma motion together with the RT equation. It employs a LTE equation of state and includes non-LTE radiative cooling in the corona and upper chromosphere, also considering thermal conduction along magnetic field lines. The electron density was computed assuming LTE ionization for all relevant species. Photoionization by hydrogen Lyman lines was not taken into account. 

The snapshot has $256 \times 128 \times 213$ grid points and a physical size of $16.6  \times  8.3 \times 5.3$ Mm. To our aims we selected a volume\footnote{The portion chosen from the data cube found in \cite{Leenaarts:2009} spans from $0$ to $5.85$ Mm in the $x$ direction, from $1$ to $6.98$ Mm in the $y$ direction and from $-0.5$ to $3.5$ Mm along the vertical.} with $5.85 \times 5.98 \times 4$ Mm ($91 \times 93 \times 191$ grid points). The snapshot has a mean magnetic field strength of $120$ G at $300$ Km, which is representative of the magnetization expected in quiet regions of the solar photosphere \citep{Trujillo-Bueno:2004, Asensio-Ramos:2014aa}.
 The hydrogen number density was not available in the supplied model. To compute it, we considered the stratifications of temperature, density and electron number density and we solved the chemical equilibrium and ionization equations for all the relevant atomic species, including hydrogen, as explained in \cite{Asensio-Ramos:2004}. Synthesis of polarization produced only by Zeeman effect in similar atmospheric models have been presented by \cite{delacruz12}.

We used the corresponding 3D NLTE Ca {\sc ii} level populations provided by Leenaarts et al. (2009) as inputs to solve the radiative transfer problem of the second kind described in \cite{ll04}, which results from considering atomic polarization and Hanle effect. Namely, it implies to solve the RT equations for the Stokes vector together with the statistical equillibrium equations (SEE) for the multipolar tensor components of the atomic density matrix $\rho^K_Q(\mathrm{J})$, with $K=0,...,2\mathrm{J}$ and $-K\leq Q \leq K$  in each energy level J. Thus, the atomic level populations ($\propto \rho^0_0(\mathrm{J})$) that were calculated with 3D RT for each level are iteratively redistributed among the corresponding magnetic energy sublevels. Essentially, it gives rise to the alignment terms $\rho^2_0(\mathrm{J})$, sourced by the radiation field anisotropy in the SEE, and to the quantum coherences $\rho^2_Q(\mathrm{J})$ (with $Q\neq0$), which can be created and destroyed by the magnetic field through the so-called magnetic kernel in the SEE (Hanle effect). %HERE%  

 The Hanle effect is also contained in the radiative transfer coefficients. Emissivities, absorptions and magneto-optical terms are the same as in \cite{manso10}. Although using populations obtained with 3D RT, we calculated the radiative transfer of the Stokes vector treating
each model column as an independent plane-parallel atmosphere, hence neglecting the effect of horizontal inhomogeneities in the plasma\footnote{This would be equivalent to the 1.5D
approximation but we prefer to reserve that terminology to the unpolarized case for avoiding confusions.}. Furthermore, we took into account that in weak magnetic fields Stokes V is due to the longitudinal Zeeman effect while Stokes Q and U are due to scattering and Hanle
  polarization, being the linear signals of the transverse Zeeman
  effect a second-order contribution. In that case the radiative
  transfer equations for Stokes Q and U can be decoupled from the RT equation
  for Stokes V. This approach is excellent for modelling the polarization in the quiet Sun, there where $B \lesssim 100$ G. Following it we obtained the linear polarization in Hanle regime and the circular polarization in Zeeman regime.

 Depolarizing elastic collisions with neutral hydrogen are taken into account \citep{Manso-Sainz:2003,manso10}, in spite of their small impact in the considered Ca{\sc ii} atomic levels. \cite{Derouich:2007aa} gives the inelastic collisional rates and the alignment transfer rates used in our calculations. A more detailed description of our synthesis computer code is presented in Appendix \ref{app:one}. Previous applications of a similar code can be found in \cite{Carlin:2013aa} and in references therein. %(Chapter \ref{cap:one}).

% Clu and Cul: excitation and de-excitation inelastic collisional rates. 
% C(2)lu and C(2) are (inelastic i think) collisional transfer rates for alignment between polarizable levels 2, 3, and 5. Derouich et al. (2007).Alternatively, Landi Degl’Innocenti & Landolfi (2004): and consider C(2) = Clup where p is just a numerical factor depending on the quantum numbers of the transition. Creo que:Born approx is applicable.
% D(K) is the ul i depolarization rate of the Kth multipole of level i due to elastic collisions with neutral hydrogen. Values given by Derouich et al. (2007)

%Finally, we add a constant microturbulent velocity $\mathrm{v_{\mu}=3.5\, km\cdot s^{-1}}$ to fit the calculated average of the $\lambda8542$ line intensity of the model with the emergent intensity provided by a solar atlas \citep{kurucz_atlas84}.
 The selected data cube contains $8463$ columns (hereafter, models or pixels in the maps). In order to decrease the computational load, we eliminated the innecessary points along the vertical direction by truncating colums in height adaptively. For each column, we estimated the heights where $\tau^{8498}_{\mathrm{cont}}=10^{3}$ (setting the lower boundary position) and $\tau^{\mathrm{K\,line}}_{\nu_0}=10^{-4}$ (setting the upper boundary position). Thus, we run the RT calculations with less grid points but without appreciable loss of accuracy in the obtention of the density matrix elements and the Stokes vector. In other words, the boundary conditions (sufficiently optically thin outside and sufficiently optically thick inside) are still fulfilled for all the spectral lines.
\begin{figure}[h!]
\centering%
\includegraphics[scale=0.78]{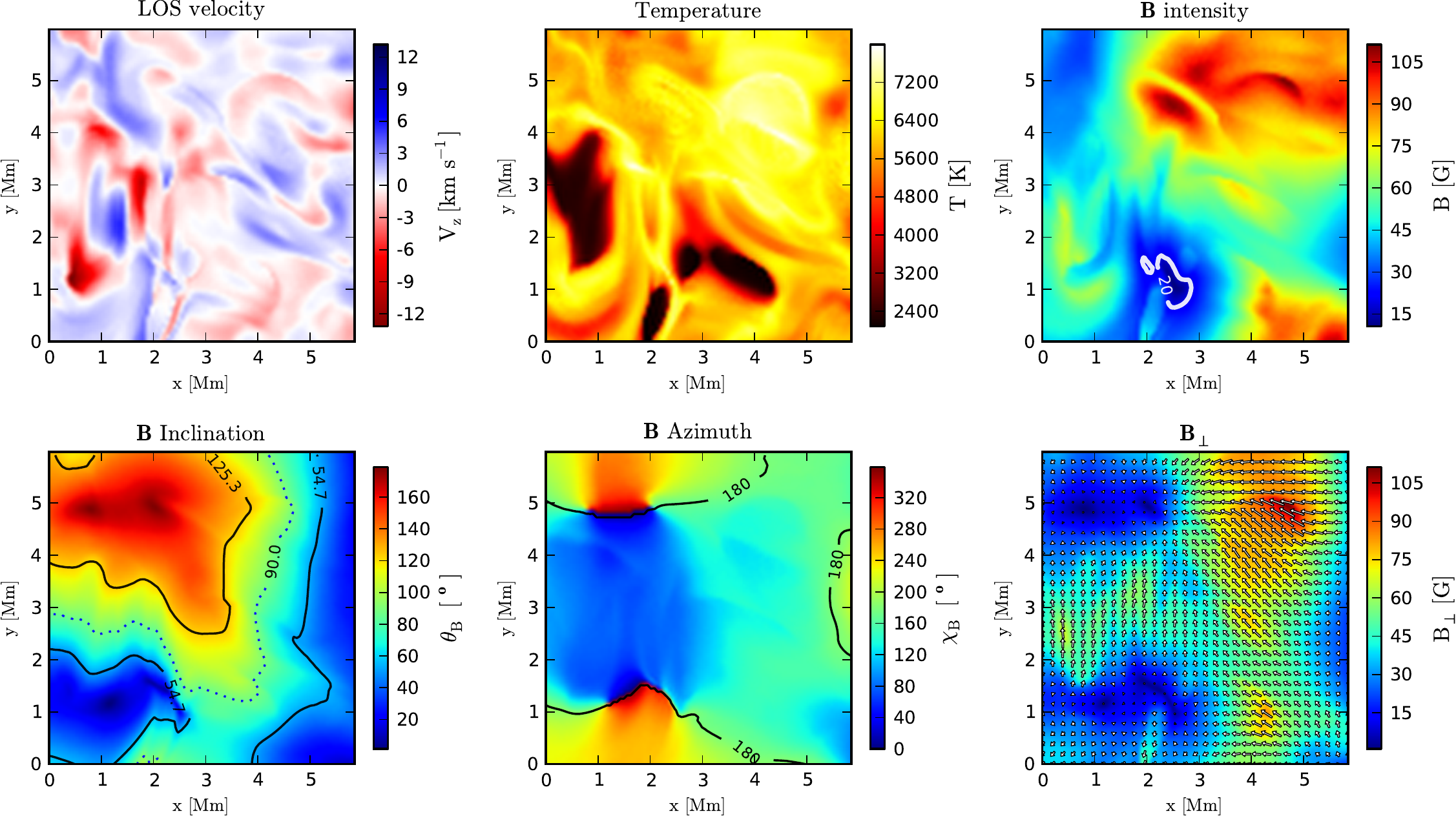} %{VTB_8542_allb.pdf}
\caption{Physical quantities at $\tau_{\nu_0}=1$ for the $8542 \,$ {\AA} line. Upper : vertical velocity, temperature and magnetic field intensity. Lower: magnetic field inclination, magnetic field azimuth and horizontal magnetic field (intensity and vector). }
\label{fig:vtb8542}
\end{figure}

To obtain the vertical limits commented above we got the iso-surfaces of optical depth by calculating, for each transition and each vertical of the data cube, the optical depth ($\tau_{\nu_0}=- \int \eta_I(z,\nu_0) dz$) at line center along rays with $\mu=1$.
% \begin{equation}
% \tau_{\nu}=- \int \eta_I(z,\nu) dz,\nonumber 
% \end{equation}
The absorption coefficient is approximated as $\eta_I \approx \,{\eta_{I}}^{\rm cont}+ \frac{h\nu}{4\pi}B_{{\ell}u}{\cal
N}_{\ell}\frac{1}{\sqrt{\pi} \Delta\nu_{D}}$,
where ${\eta_{I}}^{\rm cont}$ is the absorption coefficient for the continuum and the second addend is the line absorption coefficient at line center. $\cal N_{\ell}$ is the overall lower level population of the considered transition, $B_{\ell u}$ the Einstein coefficient for absorption, and $\Delta\nu_{D}$ is the thermal width of the line profile. This calculation is also needed to evaluate the physical magnitudes at several optical depths, specially at the main formation heights around $\tau_{\nu_0}=1$, which is useful to analyze the results of the synthesis. Examples of maps at $\tau_{\nu_0}=1$ for the Ca{\sc ii} $8542$ {\AA} line are shown in Fig. \ref{fig:vtb8542}. The corresponding heights at optical depth unity are drawn in Fig. \ref{fig:toneh}. 

The pixels having a chromospheric magnetic field that is predominantly horizontal define what we call \textit{Horizontal Field} (HF) regions, where $\mathrm{\cos{\theta_B}< 1/\sqrt{3}}$. Thus, HF regions are confined by pixels whose magnetic field lines at $\tau_{\nu_0}=1$ have the Van Vleck inclinations\footnote{These values are relevant because appear in several quantities related with the scattering polarization (e.g., radiation field anisotropy, spherical tensors for polarimetry or equations in next section).}, defined by $\mathrm{\cos{\theta_B}= 1/\sqrt{3}}$ (solid contours in bottom-left panel of Fig. \ref{fig:vtb8542}). HF areas surround the complementary \textit{Vertical Field} (VF) areas, where chromospheric magnetic field vectors are predominantly vertical. We will use this nomenclature in our analysis.

\section{The Hanle effect in forward scattering}\label{sec:fs}
The critical Hanle field $\mathrm{B_H}$ of a given atomic level is the magnetic field value in which the Hanle effect is fully operative. In an atomic transition the upper level gives the maximum critical field $\mathrm{B_H \sim A_{u{\ell}}/(1.4 \times 10^{-6} \,g_L) }$, with $\mathrm{g_L}$ the Lande factor. For the Ca{\sc ii} IR triplet lines synthetized here, the linear polarization is in the \textit{saturated Hanle regime} because the chromospheric magnetic field is significantly stronger than the critical Hanle fields
($\mathrm{B\gtrsim 5B_H}$). Considering Hanle saturation in Eqs. (7.16) of \cite{ll04} and estimating the linear polarization in the Eddington-Barbier
approximation \citep{Trujillo-Bueno:2003a}, we have obtained the following analitical expressions that describe the line-center forward-scattering polarization and help to understand our results.
 They can be applied to non-blended and sufficiently strong spectral lines forming in the weak-field and weak-anisotropy regime of quiet stellar chromospheres:
\begin{subequations}\label{eq:step6}
\begin{empheq}{align}
\frac{Q}{I} &\simeq \, -\frac{3}{4\sqrt{2}}\cdot\sin^2{\theta_B}\cdot(3\cos^2{\theta_B}-1)\cos{(2\chi_B)}\cdot\mathcal{F} \label{eq:step6a} \\
\frac{U}{I} &\simeq \,-\frac{3 }{4\sqrt{2}}\cdot\sin^2{\theta_B}\cdot(3\cos^2{\theta_B}-1) \sin{(2\chi_B)}\cdot\mathcal{F}. \label{eq:step6b}
 \end{empheq}
 \end{subequations}
\begin{figure}[b!]
        \centering
%        \begin{subfigure}{}
                \begin{center}$
                \begin{array}{cc}
                  \includegraphics[width=0.4\textwidth]{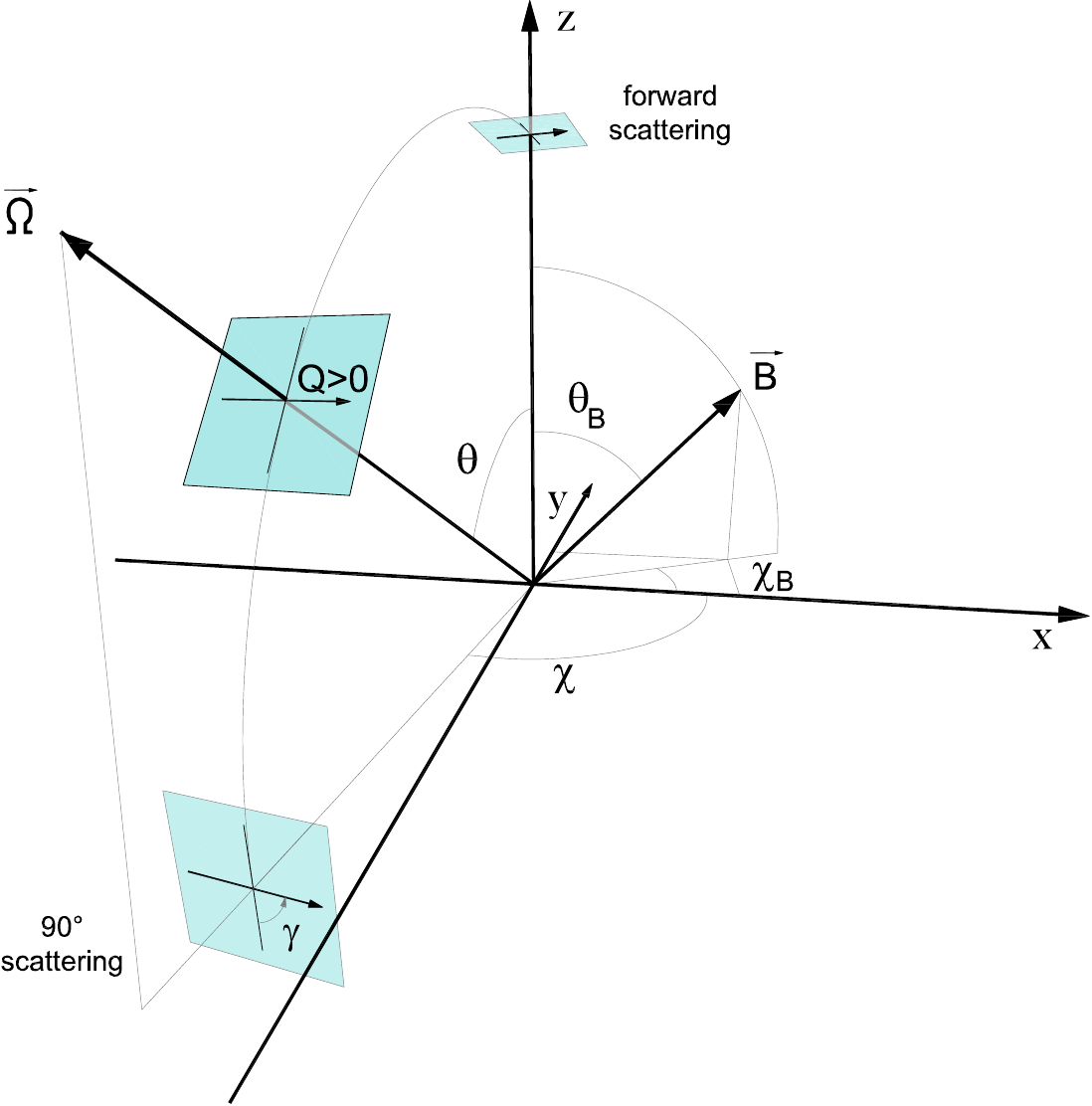}
                  \includegraphics[width=0.4\textwidth]{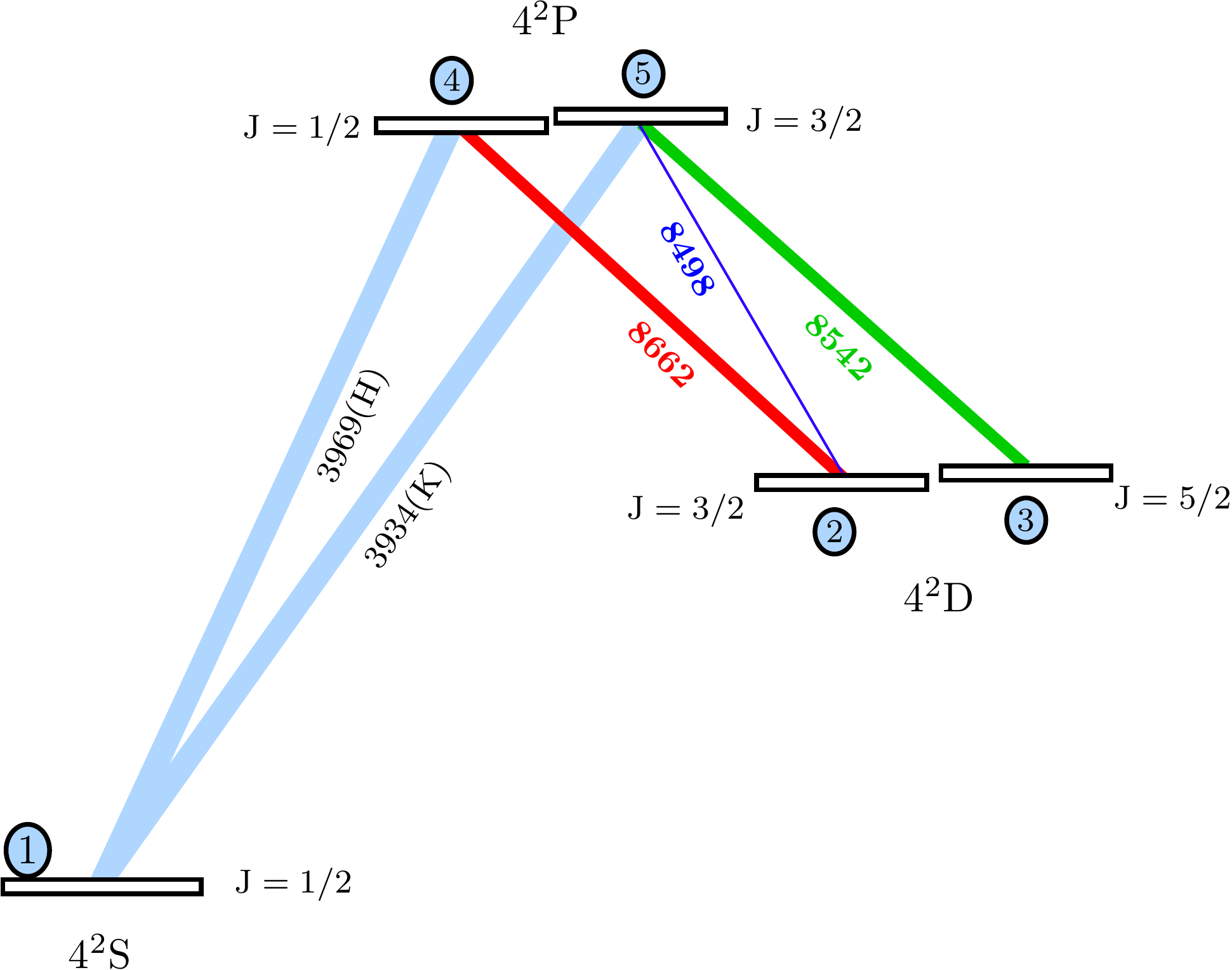} 
                \end{array}$
              \end{center}
 %       \end{subfigure}
        \caption{Left: atmospheric reference frame. The angles $\chi,\,\theta$ and $\gamma$ define the line of sight $\vec{\Omega}$ and are negative/positive in the counter/clockwise sense. Right: atomic model considered. Energy levels have been labelled from $1$ to $5$. A thicker transition line represents a larger spontaneous emission coefficient $\mathrm{A_{u\ell}}$.}\label{fig:refsys}
\end{figure}
All the terms
in the r.h.s. of Eqs. (\ref{eq:step6}) have to be evaluated at $\mathrm{\tau^{LOS}_{\nu_0}=1}$, being\footnote{We follow the standard notation for the upper and lower level atomic quantum numbers  ($\alpha_u J_u$ and $\alpha_{\ell} J_{\ell}$) and the
  polarizability coefficients $\omega^{(K)}_{J_u J_{\ell}}$ weighting the atomic polarization \citep{Landi-DeglInnocenti:1984}.} $\mathcal{F}=\omega^{(2)}_{J_u J_{\ell}} \sigma^2_0(J_u)
-\omega^{(2)}_{J_{\ell} J_u} \sigma^2_0(J_{\ell})$ the non-magnetic
contribution of the fractional atomic alignment
($\sigma^2_0=\rho^2_0/\rho^0_0$) generated in the levels of the
transition. We also call thermodynamical factor to $\mathcal{F}$ because the anisotropy and the atomic aligments modulating it indirectly depends on kinematics and thermodynamics. The angles $\chi_B$ and $\theta_B$ are the azimuth and inclination of the magnetic field vector in the reference system shown in Fig. \ref{fig:refsys}.  The LOS $\vec{\Omega}$ is defined by the angles $(\chi,\theta)=(-\pi/2,0)$ in that figure. Thus, in all our maps the direction of positive polarization given by Eqs. (\ref{eq:step6}) in the plane of the sky is parallel to the x axis for Q and inclined $45$ degrees counterclockwise from the x axis for U.
%positive-Q direction given by Eqs. (\ref{eq:step6}) in the plane of the sky is parallel to the x axis and the positive-U direction is inclined $45$ degrees counterclockwise from the x axis in all our figures. 
% \begin{figure}[h!]
%         \centering
%         \begin{subfigure}{0.2\textwidth}
%                   \includegraphics{frame3.pdf}
%         \end{subfigure}         
%         \begin{subfigure}{0.1\textwidth}
%                   \includegraphics{elevels.pdf}
%          \end{subfigure}         
%          \caption{Left: atmospheric reference frame. The angles $\chi,\,\theta$ and $\gamma$ define the line of sight $\vec{\Omega}$ and are negative/positive in the counter/clockwise sense. Right: atomic model considered. Energy levels have been labelled from $1$ to $5$. A thicker transition line represents a larger spontaneous emission coefficient $\mathrm{A_{u\ell}}$.}\label{fig:refsys}
% \end{figure}

Equations (\ref{eq:step6}) isolate in $\mathcal{F}$ the effect of the anisotropic illumination while showing that the
emergent polarization in the saturation regime of the Hanle effect does not depend on the field strength but only on its
orientation. 
The Hanle effect in forward scattering produces linearly polarized radiation being maximum along or perpendicularly to the projection of the magnetic field vector on the solar surface.
 Note that in forward scattering both Stokes parameters are essentially equivalent in their physical dependencies, having also the same maximum and minimum values. This does not occur in other lines of sight. Note also that the
axial symmetry around the solar radial (this is, a vertical magnetic field) nullifies the forward-scattering polarization. 

\section{Synthetic polarization in forward scattering.}\label{sec:profiles}
\subsection{Slit profiles.}\label{sec:profilesD}
\begin{figure}[h!]
        %~ %add desired spacing between images, e. g. ~, \quad, \qquad etc.
     %   \centering
%        \begin{subfigure}%{\textwidth}
                %\centering
                \begin{center}$
                \begin{array}{cc}
                  \includegraphics[width=0.5\textwidth]{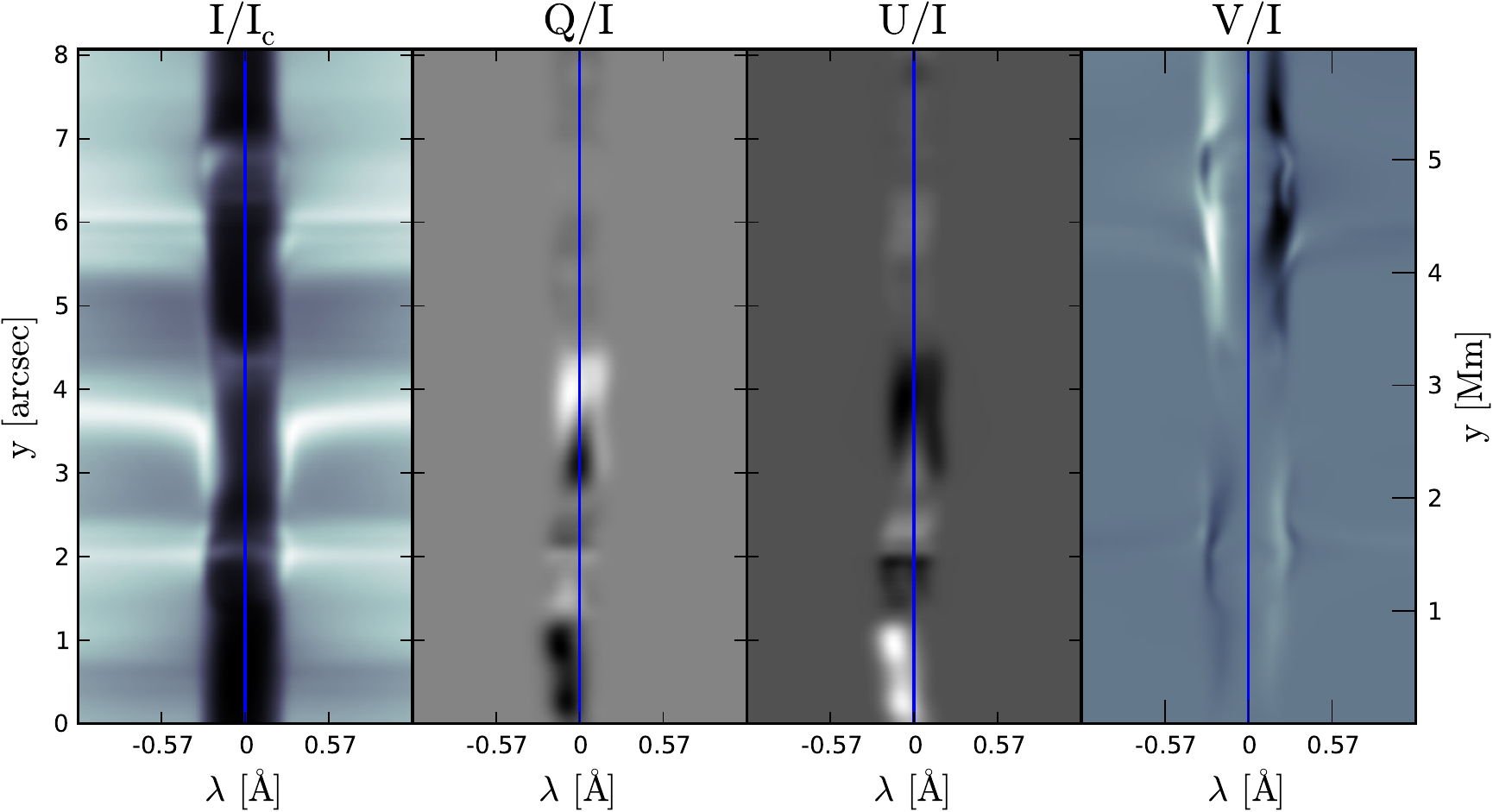} &
                  \includegraphics[width=0.5\textwidth]{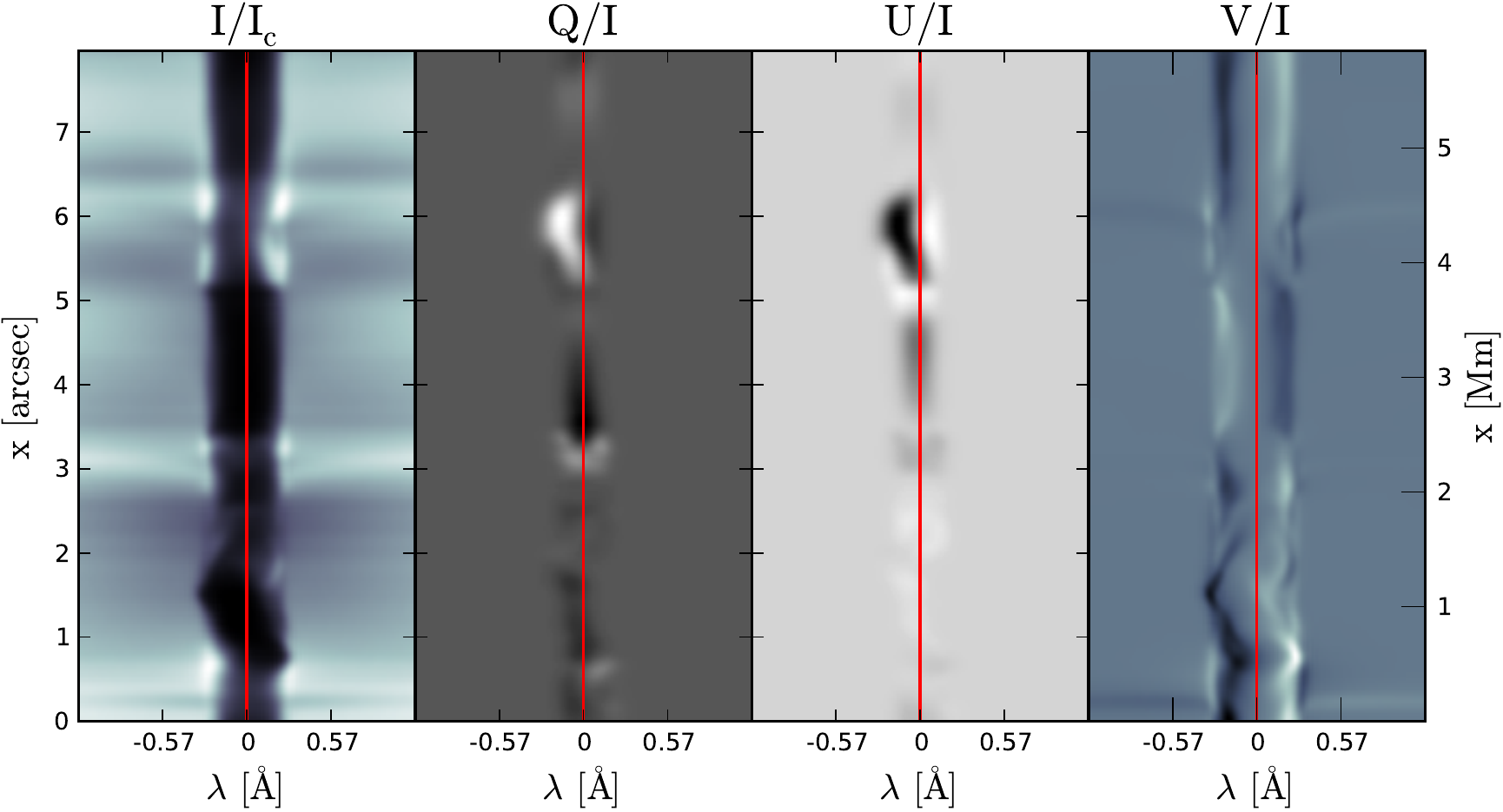} 
                \end{array}$
              \end{center}
 %       \end{subfigure}
        \caption{Examples of synthetic Stokes profiles at two simulated spectrograph slits. Left panel, blue slit: placed vertically at $x=2$ Mm. Right panel, red slit: placed horizontally at $y=2$ Mm. The left panel is analyzed in the text. Note the two different spatial scales.}
\label{fig:slitA}
\end{figure}

We have analyzed qualitatively the synthetic Stokes profiles at $8542$ {\AA} along two ficticious spectrograph's slits (Fig. \ref{fig:slitA}) while inspecting the physical situation of the  atmosphere in each point. 

First, we note that several intensity spectra in the selected slits show brightenings in the wings and/or in the core. 

On one hand, similar increased emissions in both wings of $8542$ {\AA} are also reported in solar observations \citep{Reardon:2013aa}. In our results they appear associated to photospheric heatings which, at least in the shown profiles, also correlate with photospheric compression: downward plasma is compressing sub-chromospheric layers while the low chromosphere and temperature minimum region is almost at rest. This situation sometimes coincides with a photospheric bright point in the models. In such case the magnetic field is almost vertical and shows an additional increment of strength in low layers, which produces significant Stokes V signals. That is the case at $\mathrm{y\sim2\arcsec}$  and  $\mathrm{y\sim 6\arcsec}$ in the blue slit. 

In pixels where the temperature gradient at the low chromosphere is high enough, we can also appreciate reversal peaks enclosing the intensity line-core (blue slit, $\mathrm{y\sim 3\arcsec}$) that are not necessarily correlated with wing/core emissions. These features resemble the K2 and H2 reversals of the solar H and K lines \citep{Linsky:1970aa} forming at top chromospheric layers in the Sun but, to our knowledge, do not appear in solar observations of the $8542$ and $8662$ {\AA} lines. Such discrepancy can indicate a lack of temporal and spatial resolution in observations and/or a lack of realistic temperature stratification, resolution or small-scale dynamics in the models. At least the latter option is in principle a known issue pointed out by \cite{Leenaarts:2009}. We note that introducing a height-dependent microturbulent velocity where the source function decouples from the Planck function the contributions from near layers can be mixed and the reversals be effectively supressed. The variations in the model thermal stratification can be explained accounting for other heating mechanisms \citep{Khomenko:2012aa,Martinez-Sykora:2012aa}.  

On the other hand, the weaker brightenings appearing in the intensity line core \citep[sometimes observed and catalogued as raised-core profiles; see e.g.][]{de-la-Cruz-Rodriguez:2013aa} are related to small localized increments of temperature (chromospheric heatings) in the medium-upper chromosphere (e.g., blue slit at $\mathrm{y\sim 4.3\arcsec}$). In cases with stronger temperature bumps in the formation region or with significant temperature gradients between the temperature minimum and the line formation region, the brightenings are larger and the intensity profiles are in emission (not shown here). Similar self-reversals are known to exist in the solar chromosphere \citep[e.g., ][]{Pietarila:2007aa, Judge:2010ac}.

Velocity gradients during atmospheric expansion/compression create Doppler radiative uncouplings between lower and medium-upper chromosphere that also rise the line-core intensity (Doppler brightenings). They are source of increased Q and U signals \citep{Carlin:2013aa}. The reason is that the radiation field becomes more anisotropic than in a static case, so the atomic polarization is larger \citep{carlin12}. If the velocity gradient is formed by a chromospheric shock wave, local heatings are present too. Hence, also brightenings in intensity. But this connection seems to disappear in cool expanding chromospheres. Furthermore, the scattering polarization at the disk center tends to zero inside the VF region due to the reduced magnetic field inclination avoiding the Hanle effect to act (that is why LP amplitudes in upper and lower halves of the blue-slit panel are so different). Thus, a cool chromosphere and/or the modulation produced by the magnetic field explain that a brightening in the intensity line core is not always correlated with larger  LP signals.

Examples of LP signals enhanced by vertical velocities are shown around $\mathrm{y\sim 1\arcsec}$ and $\mathrm{y\sim 4\arcsec}$ in the blue-slit panel. The essential difference between them is that $\mathrm{y\sim 4\arcsec}$ corresponds to a compressive atmosphere and $\mathrm{y\sim 1\arcsec}$ to an expansive one. Although the latter case is generally less efficient increasing the anisotropy of the radiation field, both positions have similar  LP amplitudes in this snapshot because of a larger velocity (gradient) at the chromosphere of $\mathrm{y\sim 1\arcsec}$. In both slit positions the longitudinal magnetic field strength is very weak at the chromosphere ($\mathrm{\lesssim 50 G}$) but also at the upper photosphere, which explains the absence of circular polarization. 

Scattering linear polarization signals with significant asymmetries but small Doppler shifts (e.g., Stokes Q at $\mathrm{y\sim 3.8\arcsec}$) happen in presence of steep velocity gradients but low velocities at the chromosphere ($\lesssim 3\,\mathrm{ km \,s^{-1}}$). The contrary, significant Doppler shifts and near-symmetric profiles, can happen (e.g., $\mathrm{y\sim 1\arcsec}$) when chromospheric velocities does not change their sign along the vertical but change the sign of their gradients (typical of expansive atmospheres driven by shocks). In general, the asymmetries are produced because the Doppler shifts along the LOS change the formation heights and the optical properties of the medium at different wavelengths. However, in the case of the LP profiles such RT effects are actually powered by the major contribution coming from the anisotropic illumination, which increases the vertical gradient of atomic alignments through NLTE effects in the atomic populations. In other words, vertical gradients of anisotropy are being translated to asymmetries along wavelength, which can easily surpass the assymetries produced by motions along the LOS \citep{Carlin:2013aa}. Assymetries in Stokes V and I \citep[e.g.][]{Martinez-Pillet:1990aa} are not analyzed here but naturally are also present in our results.

Several of the above-mentioned features appear together in the blue slit panel of Fig. \ref{fig:slitA} around $\mathrm{y\sim 2\arcsec}$. As commented before, it corresponds to a region over a bright point. The physical situation here is very different between the lower and the upper layers. We see reversal peaks as well as intensity brightenings in the core and the wings. Accordingly, the atmospheric analysis shows a photospheric heating by compression (downward velocities below the temperature minimum) but upward velocities in the chromosphere, which produces a significant gradient between both layers. The Hanle signal in Stokes Q and U is weak because the chromospheric magnetic field vector is near vertical but is not zero because its inclination changes significatively along height in the formation region (twisted magnetic field in the surroundings). A relatively strong \textit{vertical} magnetic field at the \textit{photosphere} and around the temperature minimum is producing a notable Stokes V signal although the chromospheric field strength is weak.

Note that the response functions to magnetic field strength in Stokes V at $8542$ {\AA} are expected to be maximum in wavelengths associated to the Stokes-V peaks and in bottom-chromosphere layers \citep{Uitenbroek:2006aa}, while the corresponding scattering polarization signals are forming substantially higher. Then, the vertical magnetic field gradients reproduced in the models give maximum amplitudes in synthetic QU-Hanle and V-Zeeman profiles that correspond to different field strengths. 

 The changes of sign in Stokes Q and U at $\mathrm{y\sim 1.3\arcsec}$ and $\mathrm{y\sim 2\arcsec}$ are due to variations in the chromospheric magnetic field azimuth (crosses through zero produced by $\chi_B$ in Eqs. (\ref{eq:step6})). Similar signatures can be easily seen in many routinary observations. 

%calentamiento por compresion de cromosfera:
%abrillantamientos en absorcion del core intensity:hay mas fotones en el core porque un dopplershifts pequeño en region de formacion (no muy caliente) desplaza una masa de plasma y la absorción es menor.
%abrillantamientos casi en emision o en emision del core intensity: lo mismo de antes pero mas exagerado, las velocidades grandes estan ahora comprimiendo zonas altas de la region de formacion, elevando la temperatura en esa region y provocando emision.
%calentamiento y magnetizacion por compresion de fotosfera:
%abrillantamientos de las alas pero un core profundo: hay compresiones por debajo del minimo de temperatura y en fotosfera. Acompañados de un incremento de campo magnetico en esas zonas bajas que se ve reflejado en grandes V/I.
%La predominancia de una emision de un tipo u otro depende del grado relativo de calentamiento fotosferico frente al cromosferico.
%Protuberantes picos en K2: calentamientos a pie de curva cromosferica, pueden ocurrir junto los anteiores caso
%Las grandes señales de polarizacion lineal son debidas a gradientes de velocidad que ocurren en zonas de campo magnetico muy horizontal en cromosfera. La diferncia entre uno y otro es que una corresponde con una compresion y el otro con una expansion, menos potente a la hora de amplificar la polarizacion lineal. La intensidad del campo es baja tanto en cromosfera como en fotosfera, lo que explica la ausencia de V.

\subsection{Polarization maps.}\label{sec:ampmaps}
\begin{figure}[t!]
\centering%
\includegraphics[width=0.98\textwidth]{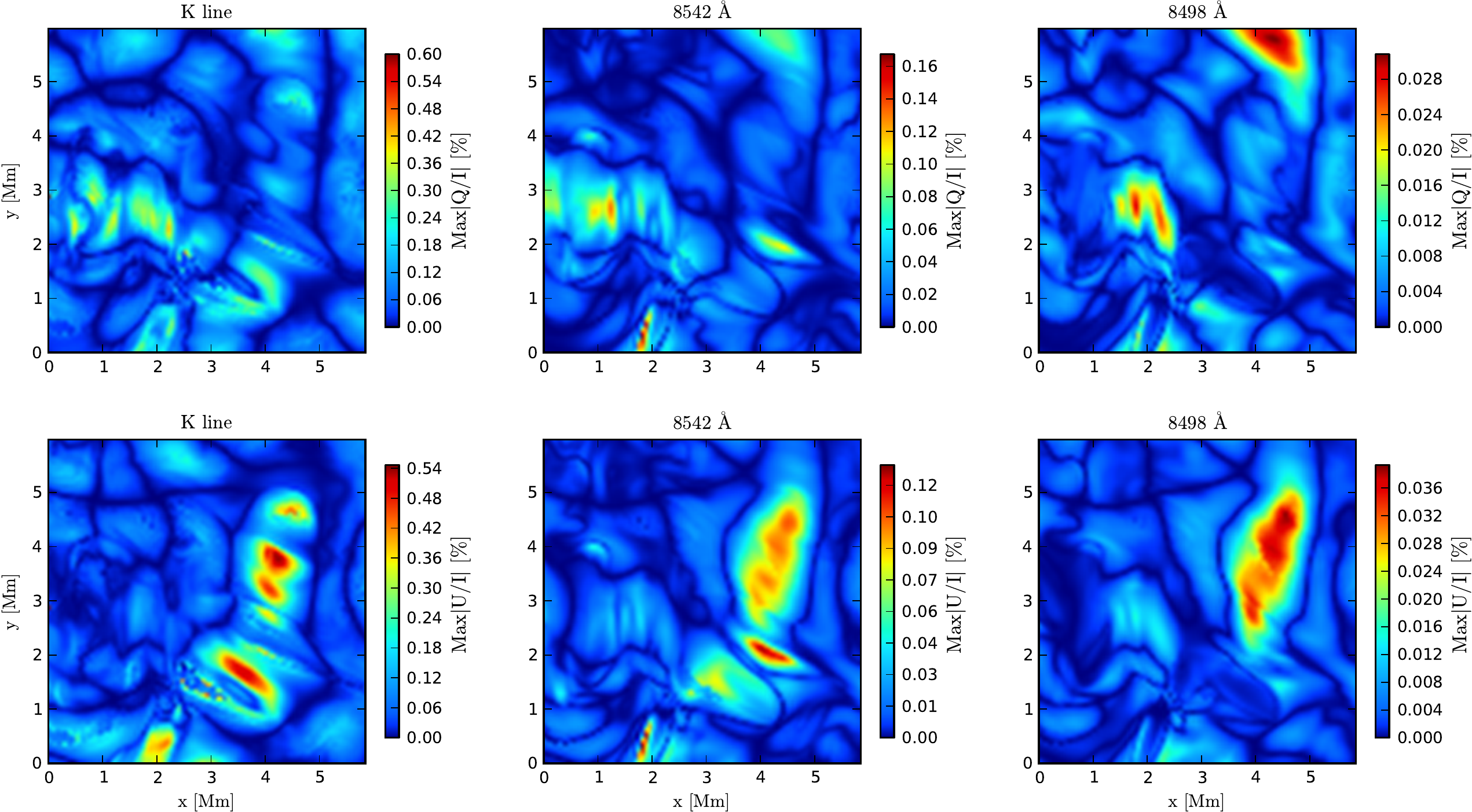}
\caption{Synthetic maps of maximum fractional polarization in Q and U for $8498$ {\AA} (right column), $8542$ {\AA} (middle column panels) and $3934$ {\AA} (K line, in the left column panels).  }
\label{fig:mapqu}
\end{figure}

Figure \ref{fig:mapqu} contains maps of maximum scattering polarization for the lines $3934$ {\AA} (Ca {\sc ii} K line), $8542$ {\AA} and $8498$ {\AA}. The quantities displayed are $100\cdot\mathrm{max|Q/I|}$ and $100\cdot\mathrm{max|U/I|}$. When these maps are made choosing other wavelengths (typically the line center) different from the ones corresponding to the maximum amplitudes, the result is a notable signal fading in pixels where chromospheric velocities are substantial. 

In general, the areas of the maps with significant LP always have a notably inclined magnetic field (HF regions). Out of such areas, the LP amplitudes are always below $1/5$ of the maximum value in the map.

In many pixels the amplitudes have the same order of magnitude as those calculated by \cite{manso10} in semi-empirical static models but, where chromospheric velocities\footnote{In general we find an association between larger velocities and larger velocity gradients, so we talk indistinctively of one or other quantity.} are above $\mathrm{\sim 5 \,km \cdot s^{-1}}$, we see relative enhancements that reach one order of magnitude in the amplitudes of the $8542$ {\AA} and $8662$ {\AA} lines with respect to static models. 
\begin{figure}[h!]
\centering%
\includegraphics[width=0.98\textwidth]{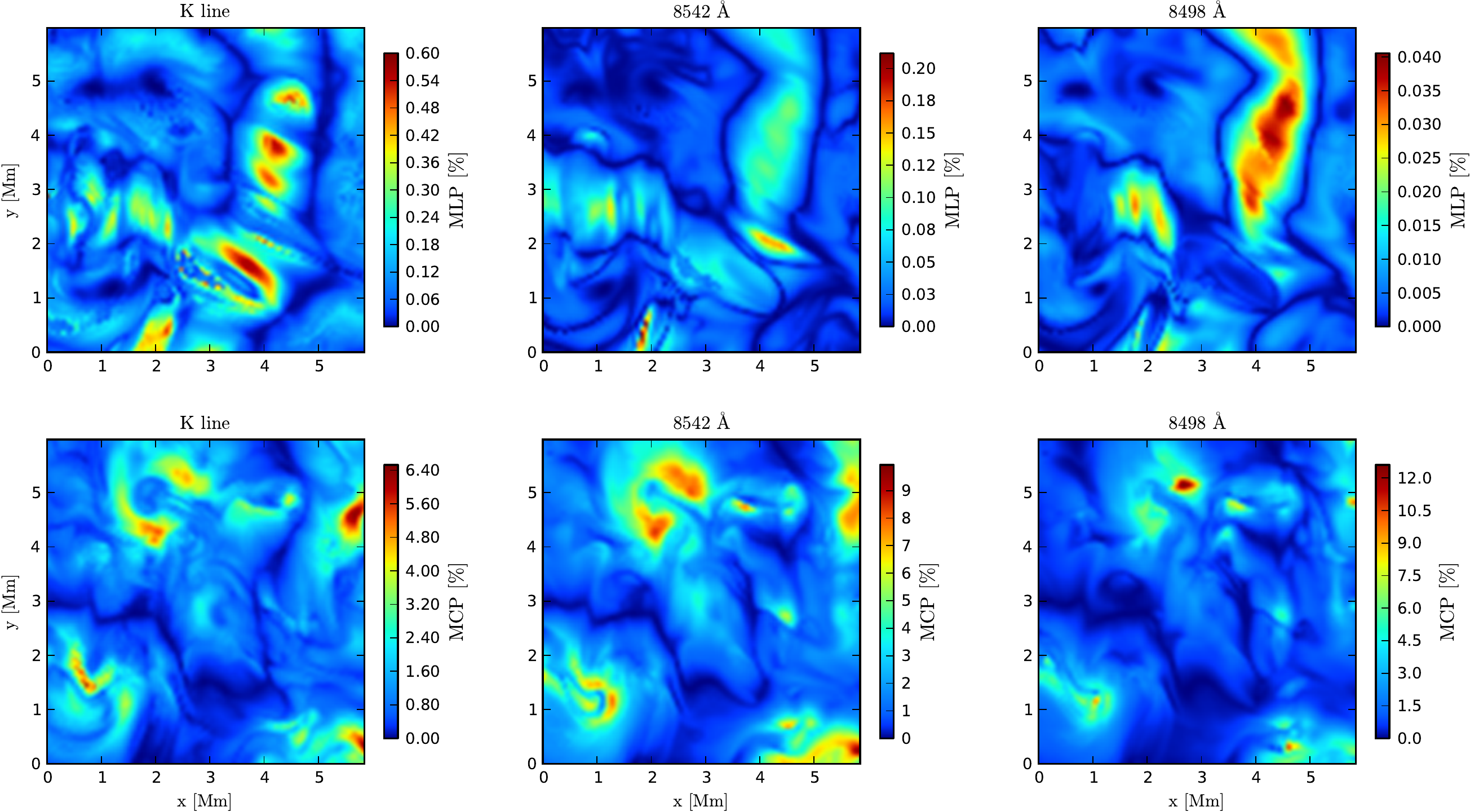}
\caption{Synthetic maps of linear and circular polarization for $8498$ {\AA} (right panels), $8542$ {\AA} (middle panels) and $3934$ {\AA} (K line, in the left panels). Top panels: maximum linear polarization amplitudes calculated as $\mathrm{MLP=\sqrt{Max|Q/I|^2+Max|U/I|^2}}$. Bottom panels: maximum of the absolute value of the fractional circular polarization (MCP).  }
\label{fig:mapmax}
\end{figure}
This is also true for the $8498$ {\AA} line only if the low chromosphere is not too cool because in that case the LP in the $8498$ {\AA} line is always insignificant (low-temperature patches in upper-middle panel of Fig. \ref{fig:vtb8542} coincide with areas of almost-zero Q and U). The largest LP amplitudes in this line appear without correlation with large velocities and are twice larger than in the FALC model. Thus, the $8498$ {\AA} anisotropy is affected by velocity gradients but seems to be dominated by temperature in these models. 

The differences between the Q and U maps are understood with Eqs. (\ref{eq:step6}). When the magnetic field existing in a region experiences a change in its azimuth, the polarization patches stand out in Q and attenuate in U or viceversa, so giving us an idea of the approximate directions of the magnetic field just by comparing Q and U regions. Patches with large Stokes Q and low Stokes U indicate that the field is chiefly oriented along some of the reference axes for Q (vertically or horizontally in the maps). If the opposite holds,  the field is then mainly oriented in directions lying at $\pm45$ degrees.

Figure \ref{fig:mapmax} shows the maximum values of the total fractional linear polarization (left panels) and fractional circular polarization (right panels) for the same three spectral lines. Pixels with negligible circular polarization are where the magnetic field is almost horizontal \textit{below} the main formation heights of the corresponding spectral line core. Such pixels are confined in patches with the largest linear polarization.

Patches with the largest linear polarization for the $8498$ {\AA} line are also the hottest in the HF region. However, for the $8542$ {\AA} line, the largest linear polarization signals are in pixels with the largest vertical velocity in the HF region. The circular polarization in the $8542$ {\AA} line is conspicuously marking the vertical magnetic field concentrations. Empirically, we find that the circular polarization amplitudes in $8542$ {\AA} at each pixel are roughly twice the corresponding ones in the K line. 

\section{Spatial signatures: Hanle Polarity Inversion Lines.}\label{sec:nullpaths}
	
In the maps of Stokes Q and U (Fig. \ref{fig:mapqu}) we observe lines where the fractional scattering polarization is zero. We call them Hanle Polarity Inversion Lines (or Hanle PILs) because they are equivalent to the polarity inversion lines of the Zeeman regime. These groove-like structures appear in the polarization maps as a consequence of the weak-field dependences that Q and U display according to Eqs. (\ref{eq:step6}). Interestingly, the HPILs encode the topology of the magnetic field pervading the solar model. In this work we restrict their description to the forward scattering geometry.

We identify three kinds of HPILs produced by three different sources that can act together. To explain them in forward scattering and Hanle saturation we use Eqs. (\ref{eq:step6}). Saturation holds easily for the $\lambda 8498$, $\lambda 8542$ and $\lambda 8662$ lines. Then, their Stokes Q and U are all described by Eqs. (\ref{eq:step6}).

The first kind of Hanle PILs is due to the inclination of the magnetic field and is explained by the term $\mathrm{sin^2{\theta_B}\cdot(3\cos^2{\theta_B}-1)}$. They appear in the same pixels for Stokes Q and U (hence also in the total linear polarization) because they are independent of the magnetic field azimuth. In the maps they can be found where the magnetic field is mostly vertical as well as in the frontiers of VF and HF regions (i.e., bordering areas with null longitudinal Zeeman polarization in Stokes V). Namely, they are where $\theta_B=  90 \pm 90^{\circ}, 90 \pm 35.27^{\circ}$, as confirmed by Eqs. (\ref{eq:step6}), so connecting pixels with such inclinations. As $\theta_B=90 \pm 35.27^{\circ}$ correspond to cases in which the magnetic field forms the Van Vleck angle with the vertical, we call Van Vleck HPILs to this first type of null polarization lines. As the magnetic field emerges in bipolar structures, the Van Vleck HPILs are continuous lines enclosing the magnetic poles. It is interesting that the mere identification of these lines means an accurate measurement of the field inclination at the main formation height of a spectral line.

 The second kind of HPILs are called \textit{azimuthal} because the corresponding spatial features depend on the magnetic field azimuth. A Hanle PIL appearing in a map of Stokes Q (or Stokes U) is of azimuthal type if it does not appear in the same place for Stokes Q than for Stokes U. In pixels defining an azimuthal HPIL in Stokes Q, the magnetic field vector is lying along the positive reference direction for Stokes U ($\pm 45^{\mathrm{o}}$ with respect to the x axis in our maps) or perpendicularly. And viceversa, the pixels defining it in Stokes U have a magnetic field vector lying along the positive reference direction for Stokes Q or perpendicularly. Following an azimuthal HPIL in the maps we connect pixels with the same magnetic field azimuth\footnote{In our definition, an azimuthal HPIL always begins and ends in an intersection of azimuthal HPILs. Thus, after such intersection, the continuation of the null line is always another azimuthal HPIL that can correspond to another azimuth.}. Note also that, when azimuthal HPILs intersect, the cross point must have a magnetic field completely vertical (so the cross point is a HPIL of the first kind). Consequently, azimuthal HPILs have a radial nature, beginning in an area of concentration of photospheric magnetic flux and ending in another one.
%As we will see in Sec. \ref{sec:b_inference}, the $90º$ and $180º$ uncertainties for the magnetic field can be avoided considering Stokes V.

Finally, a third possible origin of HPILs is a particular configuration of the anisotropy of the radiation field persisting across a region in the maps. They are \textit{thermodynamically induced }HPILs due to the $\mathcal{F}$ term in Eqs. (\ref{eq:step6}) and appear at the same time in the Q and U maps for a spectral line, just as the Van Vleck type ones, but are independent of the magnetic field inclination. To grasp some idea about the conditions in which they form, note that the thermodynamical HPILs give zero linear polarization because the non-magnetic factor $\mathcal{F}$ in Eqs. (\ref{eq:step6}) is negligible. Thus, we have 
\begin{align}
\mathcal{F}=\omega^{(2)}_{J_u J_{\ell}} \sigma^2_0(J_u)-\omega^{(2)}_{J_{\ell} J_u} \sigma^2_0(J_{\ell})=0.
\label{eq:factorf}
\end{align}
In the simplest case, given by the line $8662$ {\AA}, it yields the condition
\begin{equation}
\centering
\sigma^2_0(J_2)=\frac{\rho^2_0(J_2)}{\rho^0_0(J_2)}=0   \quad {\rm at} \quad \tau^{8662}=1%\label{eq:}
\label{eq:condicion2}
\end{equation}

How can such a condition be fulfilled? Let us suppose we have identified a thermodynamic HPIL  appearing at the same time in Q and U maps for the $8662$ {\AA} line. Equation (\ref{eq:condicion2}) is satisfied in that region\footnote{In principle, the region associated to a thermodynamical HPIL might not be a line in the map, but we still term it HPIL for consistency.} when the atomic alignment $\rho^2_0 (J_2)$ tends to zero, or when the overall population $(\varpropto \rho^0_0(J_2))$ increases too much or when both things happen at the same time. 

A larger level-2 population (see energy levels in Fig. \ref{fig:refsys}) can be achieved with an increment of temperature in the top parts of the chromosphere. Such increment strengthens the Ca {\sc ii} H  line intensity emission (forming at the top), which illuminates lower chromospheric layers from above. This extra illumination arriving at chromospheric layers immediately below (where the $8662$ {\AA} line originates) increases the population pumping from level 1 to level 4 which, in turn, produces an extra population in level 2 by spontaneous emission. Higher temperature thus means more population in higher energy levels (levels 4 and 5) and more emission (at $8662$ {\AA}) produced by electrons decaying from level 4 to level 2. Furthermore, if at the same time the formation region of the $8662$ {\AA} line is meaningfully cool (after the pass of an upward shock for instance), the absorption of electrons from level 2 to level 4 will decrease (absorption to level 5 can be neglected), so retaining the population arriving from level 4 in level 2. As the absorption between level 2 and level 4 is small the $8662$ {\AA} line cannot be polarized because its polarization can only be generated by dichroism \citep[selective absorption; ][]{manso_trujillo03a}.

%that cancels out the $\bar{J}^2_0$ component of the radiation field, which is generating alignment in level 2 from $\bar{J}^0_0$ in the SEEs
On the other hand, to have a negligible alignment in level 2 we need a formation region illuminated with a weak radiation field anisotropy. It is well-known that the radiation field anisotropy tends to zero when and where the contribution of the mainly horizontal illumination equals the contribution of the vertical one. In these models, it mainly occurs in pixels separating areas with significantly cool formation region from areas where the formation region is relatively hot. If, furthermore, such pixels does not contain significant velocity gradients, the low existing alignment will not be enhanced by kinematics. 
\begin{figure}[t!]
        %~ %add desired spacing between images, e. g. ~, \quad, \qquad etc.
     %   \centering
%        \begin{subfigure}{\textwidth}
                %\centering
                \begin{center}$
                \begin{array}{cc}
                  \includegraphics[width=0.53\textwidth]{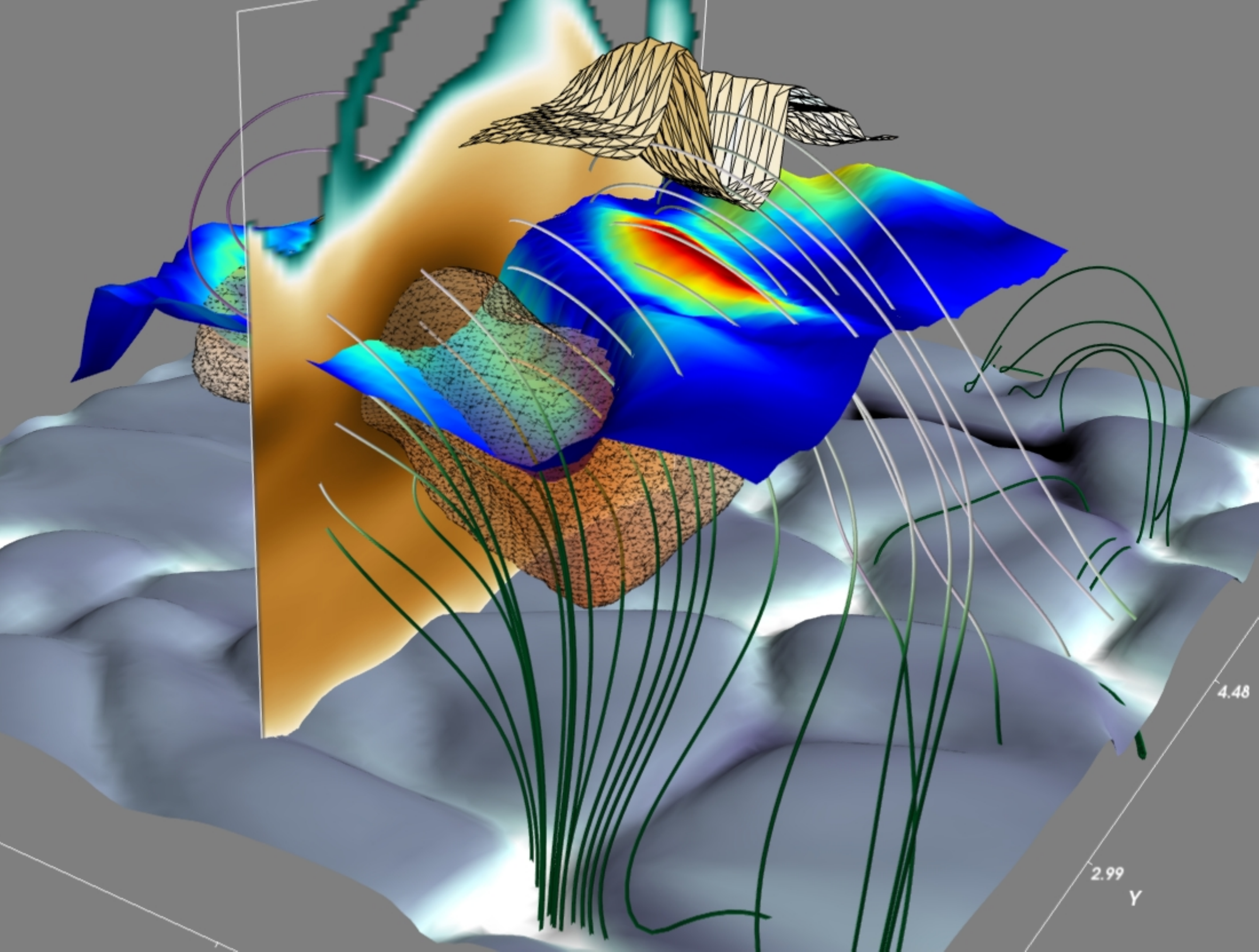} &
                  \includegraphics[width=0.45\textwidth]{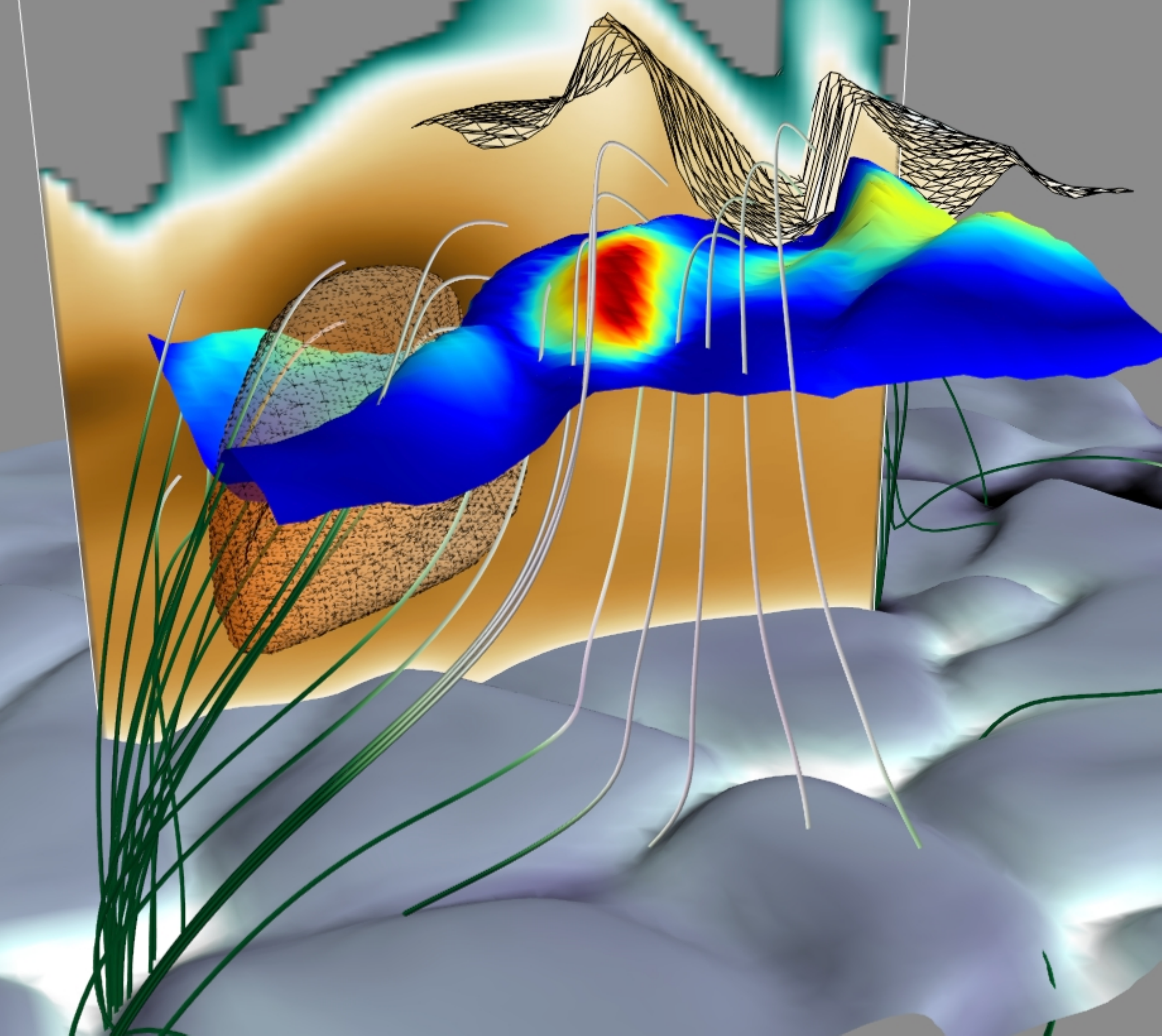} 
                \end{array}$
              \end{center}
 %       \end{subfigure}
        \caption{Visualization of a chromospheric
            ``cool bubble'' (semitransparent dotted volume). The vertical plane shows the
          temperature variation (darker brown is cooler). The
          corrugated surface sketched with polygons trace the
          heights where $\tau_{\nu_0}=1$ for the K line. The 
          coloured corrugated surface trace the heights where $\tau_{8542}=1$ and its colour
          represents the emergent Q$/$I in the $8542$ {\AA} line
          (darker blue means smaller Q$/$I). The magnetic field lines have darker colours for stronger magnetic fields. The cool bubble is tight to a depression in
            the formation layer. In the borders of the region, the linear polarization vanishes,
            so drawing a thermodynamic Hanle PIL. This is favored by the illumination of
            the K line from upper layers (polygonal surface) and by the so different kinematic and thermodynamic stratifications inside and outside the bubble. 
            }\label{fig:stereo2}
\end{figure}

The previous conditions for Eq. (\ref{eq:condicion2}) are fulfilled in the borders of cool chromospheric plasma bubbles appearing in the atmosphere. In those places the linear polarization is zero. To examine this, we first identified the location of the thermodynamical HPIL in the Stokes maps. Then, with 3D visualizations (see Fig. \ref{fig:stereo2} or their stereographic view in Appendix, Fig. \ref{fig:stereo}), we inspected the thermodynamical stratification of the verticals enclosing the cool bubble. In the interior of the volume, the chromospheric temperature is as cool as $3000$ K and the anisotropy is dominated by vertical radiation coming from above the bubble and from the photosphere. On the contrary, the line formation region outside the bubble is significantly hotter. Consequently, the horizontal radiation dominates, changing the sign of the alignment with respect to the interior of the bubble. Thus, by continuity, a line where the net alignment is zero must exist in the middle of both regions (bubble's ``frontier'') because it is positive at one side and negative at the other\footnote{There is also a correspondence with the velocities. The bubble interior is produced by an expansion cooling down the atmosphere (upward velocities) and the bubble exterior is a contraction (downward velocities). Thus, in the HPIL the velocity is almost zero or insignificant. }. That is a thermodynamical\footnote{Thermodynamical refers to temperature and velocity as drivers of the radiation field anisotropy.} Hanle PIL and it has been induced by spatial differences in the chromospheric dynamic.

A cool bubble is shown around $\mathrm{(x,y)}=(3.8,1.5)$ Mm in the temperature panel of Fig. \ref{fig:vtb8542}. The corresponding HPILs appear in Stokes Q and U maps of the $8542$ and $8662${\AA} lines, surrounding the low-temperature region. For instance, a part of the null line is connecting the points $(3.5,2)$ and $(4.5,1)$ Mm. 
The thermodynamic HPIL can be distinguished from a Van Vleck HPIL because the former is around cool patches and crosses lines of null \textit{circular} polarization.
%Ahoraora poner calculos de las configuraciones particulares de anisotropia.

 In the effective Hanle regime (this is, out of saturation), magnetic field dependences different from the ones in Eqs. (\ref{eq:step6}) can appear multiplying the density matrix components in the homologous expressions for Q and U. Thus, the approximated separation between magnetic field orientation and something similar to the thermodynamical factor could not be possible anymore. In such a case, the zeros of the linear polarization at line center can, in general, follow other analytical expressions, so defining new kinds of HPILs or nullifying the already defined here. Only when such analitical dependencies are common for all the multipolar density matrix components the corresponding HPILs would remain valid for any magnetic field regime. 

Note that the location, contrast and width of a Hanle PIL in the polarization maps gives information about the variation with height of the magnetic field along the formation region of the spectral lines. HPILs pertaining to spectral lines forming at different heights have different appearance because the magnetic field inclination changes with height. This suggests the possibility of deducing the three-dimensional topology of the chromospheric magnetic field from 2D maps of the Stokes vector in different spectral lines. A different idea is to make histograms of the mean size of the regions enclosed by the HPILs for characterizing some magnetic field parameter (helicity, inclination, azimuth) in a map. The comparison of  scattering polarization footprints between models and high-sensitivity observations would represent a very fine test to our knowledge about the quiet Sun magnetic field.

Today, the practical measurement of this structures sounds very challenging \textit{in the Ca {\sc ii} IR triplet lines}. The visual definition of HPILs in observational maps will depend on the instrumental sensitivity and resolution. Furthermore, diffuse light arriving to the detector can mask the HPILs because it diminishes the contrast between regions with and without polarization. However, some detailed calculations suggest that these structures of null polarization can be effectively distinguished in Ca{\sc ii} IR triplet lines using the forthcoming solar facilities (Solar-C, ATST, Zimpol 3+EST).

\subsection{Magnetic field intensity along Hanle PILs.}\label{sec:zee}

Though the linearly polarized Hanle signals are in Hanle saturation, we show in the following that the magnetic field intensity in the models can still be determined using the longitudinal Zeeman effect along the Van Vleck HPILs. 

In the weak-field regime, the Zeeman splitting ($\bar{g}\Delta \lambda_B$, with $\bar{g}$ the effective Land\'e factor\footnote{The Ca{\sc ii} IR triplet lines have $\bar{g}_{8498}=1.06$, $\bar{g}_{8662}=0.83$ and  $\bar{g}_{8542}=1.1$.}) is small in comparison with the thermal width ($\Delta\lambda_D$) of the line profiles:
%$\Delta\lambda_B=1.4\cdot10^{-7} \lambda^2_0[\mathrm{\AA}] B[G]$
\begin{equation}\label{eq:zeecond}
\centering
    \bar{g}\cdot \frac{\Delta\lambda_B}{\Delta\lambda_D} \, = \, \bar{g} \cdot \frac{1.4\cdot 10^{-7} \, \lambda_0[\mathrm{\AA} ] \, B[G]}{\sqrt{1.663\cdot 10^{-2} \frac{T[K]}{m[\mathrm{u.m.a.}]} +v_{\mathrm {micro}}^2[ \mathrm{km^2s^{-2}}  ] }} \ll 1,
\end{equation}
where $m=40.06$ u.m.a for calcium, $\lambda_0$ is the line center wavelength, $v_{\mathrm {micro}}$ is the microturbulent velocity and the square brackets enclose the corresponding units. In our calculations the condition (\ref{eq:zeecond}) is always fulfilled across the formation region of the spectral lines. In the absence of atomic orientation and with a constant longitudinal magnetic field component along the formation region, Stokes V in the weak-field regime is analytically approximated by:
\begin{equation}\label{eq:zeeweak}
    V(\lambda) \, = \, -\bar{g} \,\Delta\lambda_B \, \cos{\theta_B} \, \frac{\partial I(\lambda)}{\partial \lambda},
\end{equation}
where $\theta_B$ is the inclination of the magnetic field in a disk-center observation\footnote{Away from disk center, the cosine of the angle between the LOS and the magnetic field vector would not be $\cos{\theta_B}$ but $\cos{\theta_B}\cos{\theta}+\sin{\theta_B}\sin{\theta}\cos{(\chi-\chi_B)}$} with respect to the LOS and $I(\lambda)$ is the spectral profile of the emergent intensity. 
 Since $\Delta\lambda_B$ is linear in B, Stokes V is proportional to the longitudinal component of the magnetic field. Then, knowing the magnetic inclination we can get the full magnetic field intensity.

To solve for the inclination we have identified Van Vleck HPILs (Section \ref{sec:nullpaths}) in the linear polarization maps. In a hypothetical real observation such discrimination is more easily done in the maps of the \textit{total} linear polarization after integrating it around line center in order to improve the contrast between pixels in and out the HPILs. As the Van Vleck HPILs are located where $\theta_B=54.73^{\mathrm{o}}$ at $\tau_{\nu_0}=1$, the magnetic field strength B at $\tau_{\nu_0}=1$ is estimated from Eq. (\ref{eq:zeeweak}) as
\begin{equation}\label{eq:bvan}
    \mathrm{B} \,[G]\, = \, 3.71\cdot10^{12}\,\frac{|V(\lambda_{\rm core})| \, }{ \bar{g} \,\lambda^2_0 \,\left|\frac{\partial I(\lambda)}{\partial \lambda}\right|_{\rm \lambda=\lambda_{core}}},
\end{equation}
with wavelengths in Amstrongs. The ratio between the derivative of the intensity and Stokes V is wavelength dependent. For each pixel, we adaptively chose the points in a small bandwidth $\lambda_{\rm core}$ aside the line core of the Stokes V profile without taking the points in and around the peaks of the signal. We calculated the ratio appearing in Eq. (\ref{eq:bvan}) evaluating it in all the selected wavelength points and fitting it to a straight line, using its slope as the sought result. 

%Then, we chose its average value adaptively for each pixel in a small bandwidth $\lambda_{\rm core}$ aside the line core of the Stokes V profile; namely, between the spectral center equidistant from the two peaks of Stokes V and one of their inner footpoints, where the Stokes V signal starts to be significant. This defines a suficiently large portion of the profiles where the above-commented ratio is generally constant. In observations, a possible drawback can come from the effect of the noise in the central part of ``anomalously'' flat profiles related with bright points and temperature increments (see line core heatings in Fig. \ref{fig:slitA}). A similar issue has been reported in observations of raised-core profiles around active regions, where magnetic field gradients are significant \citep{de-la-Cruz-Rodriguez:2013aa}. However, outside such specific points, and specially in the quiet Sun, this should not be a problem due to the good sensitivity and signal to noise ratio in the Ca {\sc ii} $8542$ {\AA} Stokes V signal, which furthemore is translated in more wavelength points between the peaks for doing statistics. In any case, the usual technique to calculate the ratio appearing in Eq. (\ref{eq:zeeweak}) as precisely as possible is to evaluate the ratio in all the selected wavelength points and fit it to a straight line, using its slope as the sought result. 

\begin{figure}[t!]
        \centering
%        \begin{subfigure}{}
                \begin{center}$
                \begin{array}{cc}
                  \includegraphics[width=0.5\textwidth]{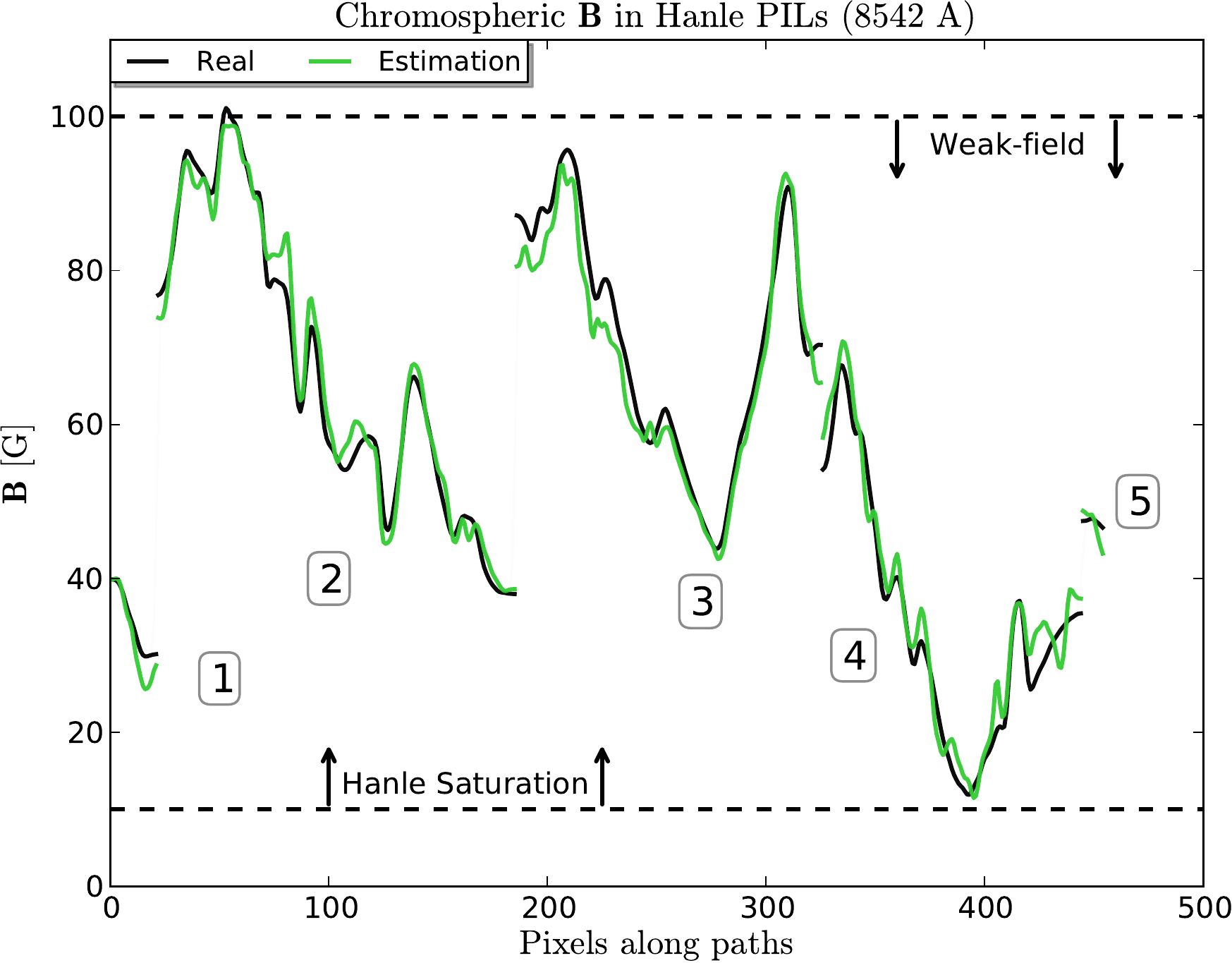}\quad\quad
                  \includegraphics[width=0.45\textwidth]{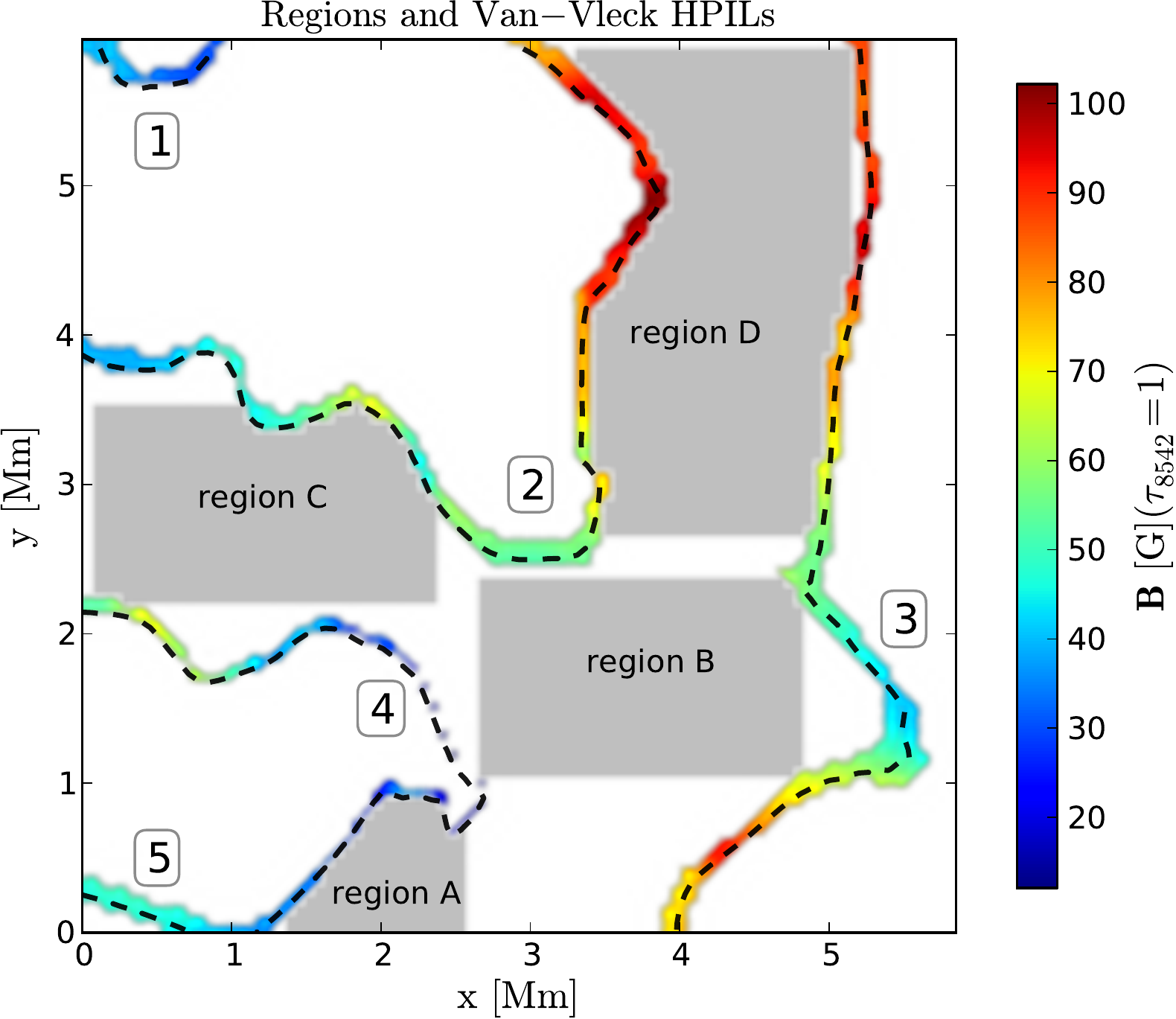} 
                \end{array}$
              \end{center}
%        \end{subfigure}
        \caption{Left: comparison between the real and the estimated chromospheric magnetic field intensity (at $\tau_{8542}=1$) in the pixels defining the Van Vleck HPILs. Right: the coloured lines are the Van Vleck HPILs obtained from Stokes Q and U, the colors representing the chromospheric magnetic field intensity. The gray regions contain the pixels considered for making the Hanle diagrams of Fig. \ref{fig:hdiag} 
          }\label{fig:bhpils}
\end{figure}

The magnetic field strenghts resulting from the calculation along the five Van Vleck HPILs found for the Ca {\sc ii} $8542$ {\AA} line are shown in left panel of Fig. \ref{fig:bhpils}. %The heights associated to each pixel can be derived from Fig() in Appendix.
Pixels where temperature and velocity gradients are larger present larger (but not large) discrepancies between estimated and real magnetic field intensities. The comparison shows that the spectral lines are formed in the weak field regime because Eq. (\ref{eq:zeeweak}) effectively allows to recover the original magnetic field in the models, despite of the small deviations due to dynamics and vertical gradients.

One could argue that the circular polarization produced by the longitudinal Zeeman effect in the Ca {\sc ii} $8542$ {\AA} line is of limited application for diagnosing the chromospheric magnetism because the response function of Stokes V is only significant from the lower chromosphere downwards \citep{Uitenbroek:2006aa}. However, we find that a correct inference of the magnetic field in the bulk of the medium chromosphere seems posible when evaluating Eq. (\ref{eq:bvan}) as explained above: including some wavelengths in the footpoints of the Stokes V lobes but avoiding the peaks. Remarkably, as those footpoints are more robust to noise than the very core they serve well for anchoring the slope of the ratio in Eq. (\ref{eq:bvan}) when is calculated with a linear fit. On the contrary, selecting wavelengths in the Stokes V peaks, the magnetic field strength delivered by Eq. (\ref{eq:bvan}) effectively correspond to deeper layers, as expected. A similar restriction to core wavelengths has been done by \cite{Woger:2009aa} when applying the Hybrid Bisector-COG method to quiet Sun observations of Ca {\sc ii} $8542$.

\section{Dynamic signatures in Hanle diagrams .}\label{sec:hanledia}
The dependency of the scattering polarization signals on the radiation field anisotropy and solar kinematics/thermodynamics, suggests the use of Hanle diagrams for studying the dynamic evolution of the chromosphere jointly with its magnetic field. The development of diagnostic metrics based upon such diagrams requires to understand the signatures produced by the chromospheric dynamic events. Next, we use ``dynamic'' Hanle diagrams to elaborate on this purpose and to establish basis for deeper developments. 

\begin{figure}[t!]
        \centering
%        \begin{subfigure}{}
                \begin{center}$
                \begin{array}{cc}
                  \includegraphics[width=\textwidth]{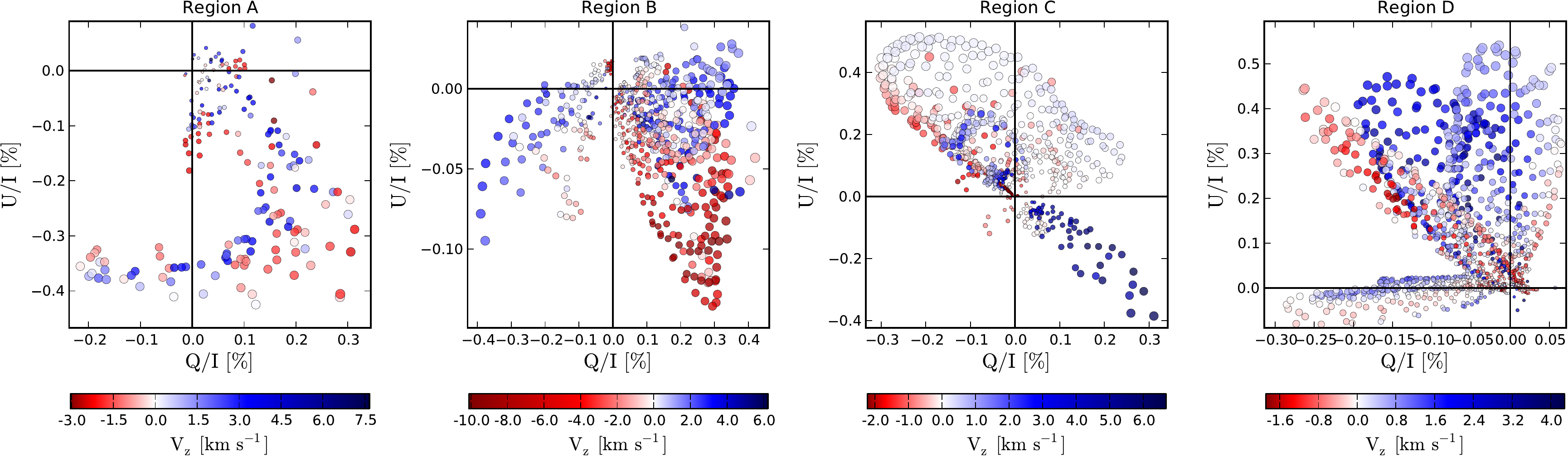}\\
                  \includegraphics[width=\textwidth]{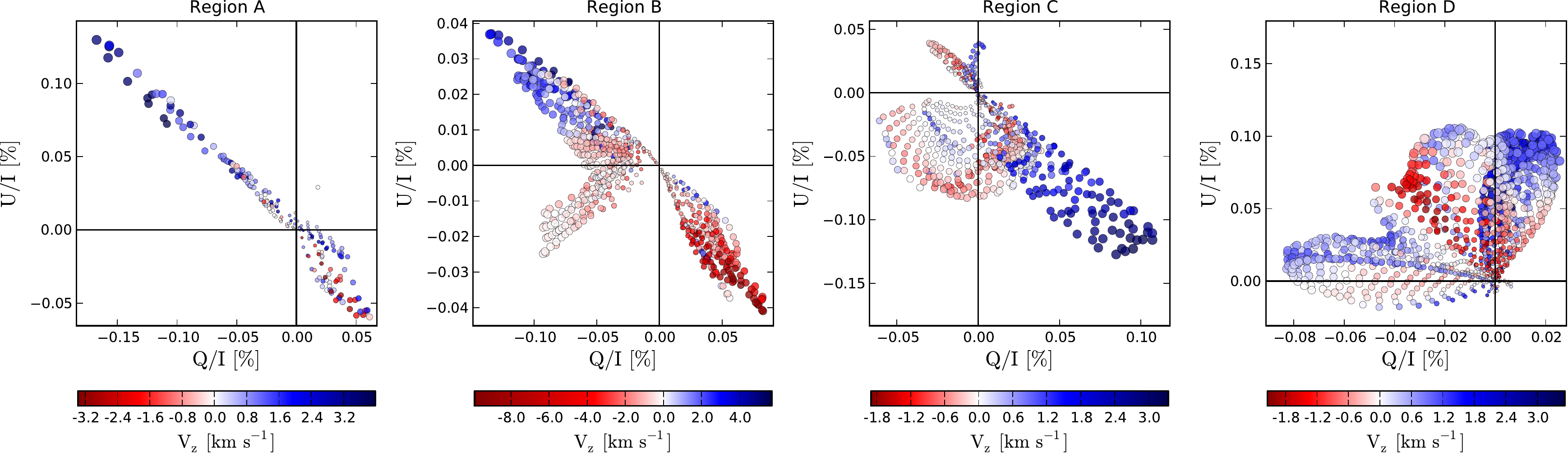} \\
                \end{array}$
              \end{center}
 %       \end{subfigure}
        \caption{ Forward scattering Hanle diagrams in the selected regions of Fig. \ref{fig:bhpils} for the $3934$ {\AA} and the $8542$ {\AA} lines  (top and bottom panels, respectively). Each small circle is slightly semitransparent and represents a pixel in the corresponding region. The color encodes the value of the vertical velocity (at the height of formation of the corresponding line) and the size of each circle is proportional to the distance to the axes origin.
          }\label{fig:hdiag}
\end{figure}

We studied the synthetic Hanle diagrams for the $8542$ {\AA} and Ca {\sc ii} K lines (Fig. \ref{fig:hdiag}, lower and upper panels, respectively) in four regions of our maps. For comparison, a suitable reference for the $8542$ {\AA} Hanle diagrams in (non-dynamic) FALC models and forward-scattering can be found in Figures 14 and 15 of \cite{manso10}. The four spatial regions selected in our maps are labelled as A, B, C, and D in the right panel of Fig. \ref{fig:bhpils}, being all of them in the HF region. The small Region A has an expanding chromosphere with significant velocity. The region B is a cool area with patches in different states of motion: static, upward and downward (having strong downflows). The region C contains a cool bubble with downward velocities and an elongated portion with expanding hot plasma. Finally, region D is the largest, hottest and with strongest magnetic field, but has negligible kinematics.
Each small circle in the Hanle diagrams corresponds to one pixel of the spatial map, and their colours encode the values of the vertical velocity at $\tau_{\nu_0}=1$. The correspondence between the position of each pixel in the map and the position of each pixel in the Hanle diagram can be easily followed in synthetic (and observational) maps. Different small-scale structures of the quiet Sun can thus be studied as different curves and shapes in the Hanle diagram, with the advantage that the axes of the Hanle diagram have associated a physical meaning encoding anisotropy, magnetic field and dynamics. 

Using the approximate Eqs. (\ref{eq:step6}) we can understand the Hanle diagrams for the $8542$ {\AA} line (Fig.  \ref{fig:hdiag}) as curves in polar coordinates. The corresponding radius r of a given point depends on the inclination of the magnetic field and on the thermodynamic factor $\rm{\mathcal{F}}$, but with no dependence on the field strength because the IR triplet is Hanle saturated:
\begin{subequations}\label{eq:polar_diag}
  \begin{align}
    \mathrm{\left(\frac{Q}{I}\right)} &=\,  r(\mathcal{F},\theta_B) \cdot  \cos{2\chi_B},\label{eq:param_a}
\displaybreak[0] \\
    \mathrm{\left(\frac{U}{I}\right)} &= \, r(\mathcal{F},\theta_B) \cdot  \sin{2\chi_B}.\label{eq:param_b}
  \end{align}
\end{subequations}

As the points drawn in the lower row of Fig. \ref{fig:hdiag} for the $8542$ {\AA} line correspond to the same pixels in the spatial maps than the points in the upper row, the whole figure illustrates the effect of changing from a \textit{medium-chromosphere} line to a \textit{top-chromopshere} line in the (Q,U) space. In the case of the K line, the assumption of saturation in the Hanle effect is in principle not valid; on one hand because this line forms at top chromospheric layers, where the magnetic field intensity is  reduced (compared to lower layers); but, overall, because the intrinsic critic Hanle field of the upper level of the K line is large enough, tipical of resonant lines with large emission coefficients $A_{ul}$. In that case, the polar radius of a given point in the Hanle diagram depends also on the magnetic field strength.

However, there is no observational rule to know which areas of a Hanle polarization map are effectively saturated. Our analysis suggest that the signature of the sensitivity to the magnetic field strength in the K line appears in Fig. \ref{fig:hdiag} (lower row panels) as a general absence of order when connecting adjacent points. On the contrary, a close inspection to the $8542$ {\AA} diagrams shows that their points are \textit{generally} aligned following organized trajectories. 
In general, those trajectories can be fit without needing changes in the magnetic field strength, but just considering continuous variations of the magnetic field orientation across the selected spatial regions in the maps. 

There are however notable exceptions in the $8542$ {\AA} diagrams that suggest a clear lack of order (examples are the darkest red points in regions B and D). The source of such exceptions in the $8542$ {\AA} diagrams is dynamics: modulation of the scattering polarization signals by velocity and temperature shocks crossing the chromosphere. As the horizontal variations in temperature and velocity are comparatively larger and mostly occur in shorter spatial scales than the ones in magnetic field orientation, the variations in the radius r between adjacent points in the diagrams can be significantly enlarged while their correspoding azimuthal displacement cannot.
 Thus, the radial excursions adding some disorder to the diagrams of the $8542$ {\AA} line are typical signatures of dynamic effects. As sudden radial variations can be distinguished from azimuthal ones, we have here a possible proxie to discriminate between the action of the magnetic field and the kinematic.

Although the dynamic modulation is also present in the K line core, the effect is not so strong because the ratio between the Doppler shifts and the thermal broadening of the absortion profiles in the top chromosphere is not as large as in the formation region of the $8542$ {\AA} line, in such a way that the action of the velocity gradients on the anisotropy of the radiation field is reduced \citep{carlin12}. Furthermore, the action of the magnetic intensity on the K line polarization could also be to spread out the points in azimuth, so contributing to dilute the strong radial excursions. We cannot confirm this from Eqs. (\ref{eq:step6}) because they are valid for Hanle-saturated lines.

A similar analysis can be carried out using local temperature instead of velocity. We find  correlations between points with larger temperatures, larger velocities and larger Q and U signals for the $8542$ {\AA} line. All suggest the picture of a quiet chromosphere where the Hanle-saturated signals of the $8542$ {\AA} line mapping the magnetic field orientation are abruptly modified by the dynamic effects associated with the propagation of shocks.
  
The atmospheric model here employed has larger resolution and reduced kinematics in comparison with the real observed Sun, which suggest that the dynamic effects above commented could be significantly stronger. The improvements in resolution, both in synthetic models and in observational maps, will transform the sub-pixel (now microturbulent) kinematic into macroscopic velocities, so making the effects presented in this paper increasingly more evident and relevant for reproducing the observations.
\newpage
% %%%%%%%%%%%%%%%%%%%%%%%%%%%%%%%%%%%%%%%%
\section{Conclusions}
% %%%%%%%%%%%%%%%%%%%%%%%%%%%%%%%%%%%%%%%%

A better understanding of the forward-scattering polarization signals requires two-dimensional spectropolarimetry for measuring polarization maps instead of slit profiles. Magnetic field diagnosis at (or near) forward scattering shows several unexploited advantages: first, although the forward-scattering signals have lower amplitude because the level of polarization produced by solar curvature is naturally removed at disk center, the number of collected photons is significantly increased with respect to the limb (limb darkening); second, the geometry allows the creation of Stokes vector maps whose pixels are directly adscribed to a single vertical stratification in the atmosphere, which minimizes second-order RT effects expected from horizontal inhomogeneties and avoids mixing of horizontal structures along the LOS; third, in contraposition to slit-like observations, the spatial continuity in the maps brings an important help to understand the magnetic topology and to discriminate the magnetic and kinematic contributions to the linear polarization signals; and fourth, partial redistribution effects in the scattering process are minimized at disk center, which supress polarized emission in the spectral wings. As the effects of the solar curvature and PRD in the linear polarization are avoided, ``purer'' profiles remain at disk center. They are driven by magnetic field (Hanle effect) and solar dynamics (through the radiation field anisotropy).   %%%%HEREE

Thus, the fundamental point is the discrimination between Hanle and dynamic effects, because both modulate the line-core polarization (independently, in principle). Such discrimination is possible by understanding the variability and the characteristics of the polarization signals in wavelength \citep{carlin12}, time \citep[][]{Carlin:2013aa} and space (this paper) from realistic atmospheric models.
Analyzing jointly the Stokes vector spectrum, the spatial information in the polarization maps and the atmosphere model, we have outlined the expected behavior of the chromospheric Ca {\sc ii} lines in a ``realistic'' quiet Sun context.
We conclude that the combination of the forward-scattering Hanle effect with temperature and velocity dynamics should produce characteristic spatial patterns in the linear polarization signals of spectral lines forming at the solar chromosphere. When these signals are synthetized in MHD models we find that the Hanle effect operates in the saturation regime for all the lines in the Ca{\sc ii} triplet. Thus, with the aid of analytical formulae derived for this case, we have introduced the concept of Hanle Polarity Inversion Lines, defining and explaining three types: azimuthal, Van Vleck and thermodynamical HPILs. The so-called thermodynamical HPILs can offer new diagnostic metrics to capture very specific properties of the dynamical state of the atmosphere, like the ones created around cool chromospheric volumes of plasma. The other kinds of Hanle PILs found in our synthetic Ca {\sc ii} IR triplet signals encode the orientation of the chromospheric magnetic field, representing a precious magnetic fingerprint of the chromosphere. 

Using the Stokes V profiles we have shown how the Van Vleck HPILs could serve to infer the chromospheric magnetic field strength along themselves. Hence, the detection of HPILs is not only a direct measurement of the magnetic field orientation but also an interesting constraint for the magnetic field strength. This discussion suggests a question: having a high-resolution and high-sensitivity polarization map at disk center, can we get a precise mapping of the magnetic field using the spatial locations where the polarization amplitudes \textit{cancel out}? %%HERE

Our results should be tested observationally with chromospheric spectral lines and instrumentation providing good signal to noise ratios with enough spatial resolution. The extension of this work to the full 3D radiative transfer case will quantify whether the horizontal inhomogeneities of the plasma can mask the location of the HPILs or smooth their contrast significatively.  

In this paper we have also calculated synthetic Hanle diagrams characterizing a dynamic chromosphere in presence of magnetic field. The spatial and temporal evolution of the points in such diagrams translates the physical problem of the diagnostic to a more suitable representation and suggests that the variations of magnetic field and temperature/velocity could be studied and understood jointly. On one hand, the signatures of an amplitude modulation produced by vertical  gradients of velocity and temperature are abrupt radial excursions of the points in the Hanle diagram. On the other, points that follow ordered curved trajectories in the diagramas seem to indicate Hanle saturation in the corresponding region of the spatial maps. Their variations are guided by magnetic field azimuth and inclination. On the contrary, when the Hanle effect is not saturated the polarization points cannot be easily adscribed to a parametric curve because the action of the magnetic fied intensity spread them out in a more chaotic way across the (Q,U) space. Provided that the noise uncertainties do not exceed a certain threshold, this idea can serve to easily diagnose Hanle saturation in regions of the solar surface.%This indirectly implies an estimation of a magnetic field strenght threshold.

 Our results are expected to be suitable for more chromospheric lines.
In the next papers of this serie we will show additional calculations that complement these results.

\acknowledgments 
We thank to Javier Trujillo Bueno and Rafael Manso Sainz for helpful discussions and advices and to Mats Carlsson and J. Leenaarts for providing us with the atmospheric models. ESC gratefully aknowledge the financial support given by the Istituto Ricerche Solari Locarno (IRSOL) through the State Secretariat for Education, Research
and Innovation (SERI, Canton Ticino), the COST project C12.0084, the
municipalities affiliated to CISL and the Aldo e Cele Dacc\`o Foundation as well as by the Instituto de Astrof\'isica de Canarias through the Resident Astrophysicist fellowship. Financial support by the Spanish Ministry of Economy and Competitiveness 
through projects AYA2010--18029 (Solar Magnetism and Astrophysical Spectropolarimetry) and Consolider-Ingenio 2010 CSD2009-00038 are also gratefully acknowledged. AAR acknowledges financial support through the Ram\'on y Cajal fellowships.

\appendix

\section{The RT code}\label{app:one}
Traviata is a multilevel RT code for the synthesis of the spectral line
polarization resulting from atomic polarization and the Hanle effect in weakly
magnetized stellar atmospheres
\citep{Manso-Sainz:2003, manso10}. Esentially, it solves the RT problem of the second kind in the framework of the quantum theory of spectral line polarization under the flat-spectrum approximation. Quantum coherences and population imbalances between magnetic energy sublevels of any given J level are considered. 
We have added new functionalities to 
such code. Namely:
\begin{enumerate}
\item  Effect of vertical (non-relativistic) velocities in the RT and in the Statistical Equillibrium Equations. 

The calculation is posed and solved in an external
observer's reference frame, which all the motions in the plasma are related to. The
 absorption, emission and dispersion coefficients depend on position, 
  frequency, spectral
 transition and, due to the velocity, on the inclination of the
 rays of the angular quadrature. Upward velocities gives
 blue shifts in the spectral profiles.

\item Adaptative numerical grids. 
 
For each column of the model, a symmetric frequency axis is created, having a variable adaptative resolution to correctly sample the Doppler-induced features along the Stokes profiles. Initially, it is calculated for each vertical considering the existing maximum velocity and remains fixed thereafter. Angular and spatial numerical grids are heavily restricted by the presence of velocities due to the maximum Doppler shift at each column. For
 a good spectral sampling, the change in Doppler velocity
 $\Delta(\mu_k V_i)=\Delta\mu_k \cdot V_i+\Delta V_i \cdot\mu_k$ in
 the spatial and angular grid has to be
  small enough (i.e., of the order of half Doppler width)
  between adjacent points along the same ray $k$ and also between adjacent points along the same height $i$. On one hand, this constraints the angular grid to fulfill that $ |\Delta \mu_{\rm max} V_{\rm max}| \lesssim \frac{1}{2}$,  where $\Delta \mu_{\rm max}$ gives the maximum allowed angular step
 in the quadrature. On the other hand, the spatial grid is constrained by $\Delta V_{i}=|V_{i}-V_{i-1}|\lesssim 1/2$, which limits the maximum Doppler velocity gradient between two adjacent layers $i$ and $i-1$ in the atmosphere.

\item Variation of all the quantities with height. 

\item  Calculation of Stokes V in Zeeman regime. 

The code uses the total level populations to solve the Zeeman
  transfer equation for I and V along the observer LOS:

\begin{equation} \label{eq:zeerte}
\frac{\rm{d}}{\rm{ds}}
\left( \begin{array}{c}
 I \\ 
 V
\end{array} \right) 
=
\left( \begin{array}{c}
\epsilon_I \\ 
 \epsilon_V
\end{array} \right) 
-
\left( \begin{array}{cc}
 \eta_I  & \eta_V \\ 
 \eta_V  & \eta_I
\end{array} \right) 
\left( \begin{array}{c}
 I \\ 
 V
\end{array} \right) .
\end{equation}
The dispersion (magneto-optical) terms are neglected when calculating Stokes V because 
their contributions are insignificant in
weakly magnetized atmospheres for this Stokes parameter. To solve the system of equations we transform it in two independent equations:
\begin{equation} \label{eq:zeerteplus}
\frac{\rm{d}}{\rm{ds}}
\left( \begin{array}{c}
 I^{+} \\ 
 I^{-}
\end{array} \right) 
=
\left( \begin{array}{c}
\epsilon^{+} \\ 
 \epsilon^{-}
\end{array} \right) 
-
\left( \begin{array}{c}
 \eta^{+} I^{+} \\ 
 \eta^{-} I^{-}
\end{array} \right),
\end{equation}
creating the variables $I^{\pm}=I \pm V$, $\epsilon^{\pm}=\epsilon_I \pm
\epsilon_V$ and $\eta^{\pm}=\eta_I \pm
\eta_V$. Both equations are solved along the LOS using a parabolic
short-characteristic method to get the emergent $I^{+}$ and $I^{-}$
profiles for each frequency and spectral transition. The solution for
the original
variables is then obtained with $I=(I^{+}+I^{-})/2$ and
$V=(I^{+}-I^{-})/2$. 

 The Stokes V line radiative coefficients (absorption and emission terms) are calculated without atomic orientation, following the standard expressions for the longitudinal Zeeman effect \citep[e.g., ][]{ll04,Stenflo:1994}. Namely, the line emissivity is 
\begin{equation}\label{eq:ev}
\centering 
\epsilon^l_{V} (\nu, \vec{\Omega}) = (h\nu/4\pi)A_{u{\ell}}N_u \phi_V(\nu,\vec{\Omega}),
 \end{equation}
with
\begin{equation}\label{eq:zeeeq}
\phi_{V} (\nu, \vec{\Omega}) = {\displaystyle 
\frac{1}{2}\left[\phi_1 - \phi_{-1}\right] \cos{\hat{\theta}} }
 \end{equation}
and
\begin{equation}\label{eq:phicomp}
\centering 
\phi_q =\sum_{M_{\ell}M_u} 3 \left(
\begin{array}{ccc}
J_u & J_{\ell} & 1 \\
-M_u & M_{\ell} & -q
\end{array}
\right)^2\phi(\nu-\nu_{J_u M_u, J_{\ell} M_{\ell}}) \qquad (q=-1,0,-1)
\end{equation}
describing the superposition of Zeeman components $\phi(\nu-\nu_{J_u
  M_u, J_{\ell} M_{\ell}})$ which are evaluated around its corresponding Zeeman frequency $\nu_{J_u  M_u, J_{\ell} M_{\ell}} = \nu_0 + \nu_L (g_u M_u - g_{\ell} M_{\ell})$. For each transition $u \leftrightarrow {\ell}$, $A_{u{\ell}}$ is the Einstein coefficient for spontaneous emission and $N_{ u}$ is the total population of the upper level per unit volume.
Note also that $\hat{\theta}$ is the direct angle between the local magnetic field
vector and the direction $\vec{\Omega}$ of a given ray light passing through
  the considered plasma element. Working in the reference frame of the solar vertical, this angle can be calculated
  locally at every point in the atmosphere with:
\begin{equation}\label{eq:v_angle}
\centering
\cos{\hat{\theta}}= \cos{\theta}\cos{\theta_B}+\sin{\theta}\sin{\theta_B} \cos{(\chi_B-\chi)} ,
\end{equation}
being then $\theta$ and $\chi$ the angles defining the direction of the ray in such reference system.

\item Automatization and parallelization for working with large atmospheric models.
  The processing of the atmospheric model includes interpolation of the physical magnitudes in grid intervals with large gradients, initialization with atomic populations got in near positions in the dataset and management and re-processing of models with convergence problems, which have to be recomputed with other
  interpolation grid, with other parameter settings or with other
  initializing populations. To accelerate the convergence, we have
implemented an OpenMP parallelization on the radiative transfer loop.
\end{enumerate}

As a result the code can treat large atmospheric datasets considering vertical gradients in all the physical quantities and obtaining the LP
  signals produced by scattering and Hanle effect and the circular polarization produced by Zeeman effect.

\renewcommand\thefigure{\thesection.\arabic{figure}}    
\section{Additional figures}
\setcounter{figure}{0}  

\begin{figure}[h!]
\centering%
\includegraphics[scale=0.7]{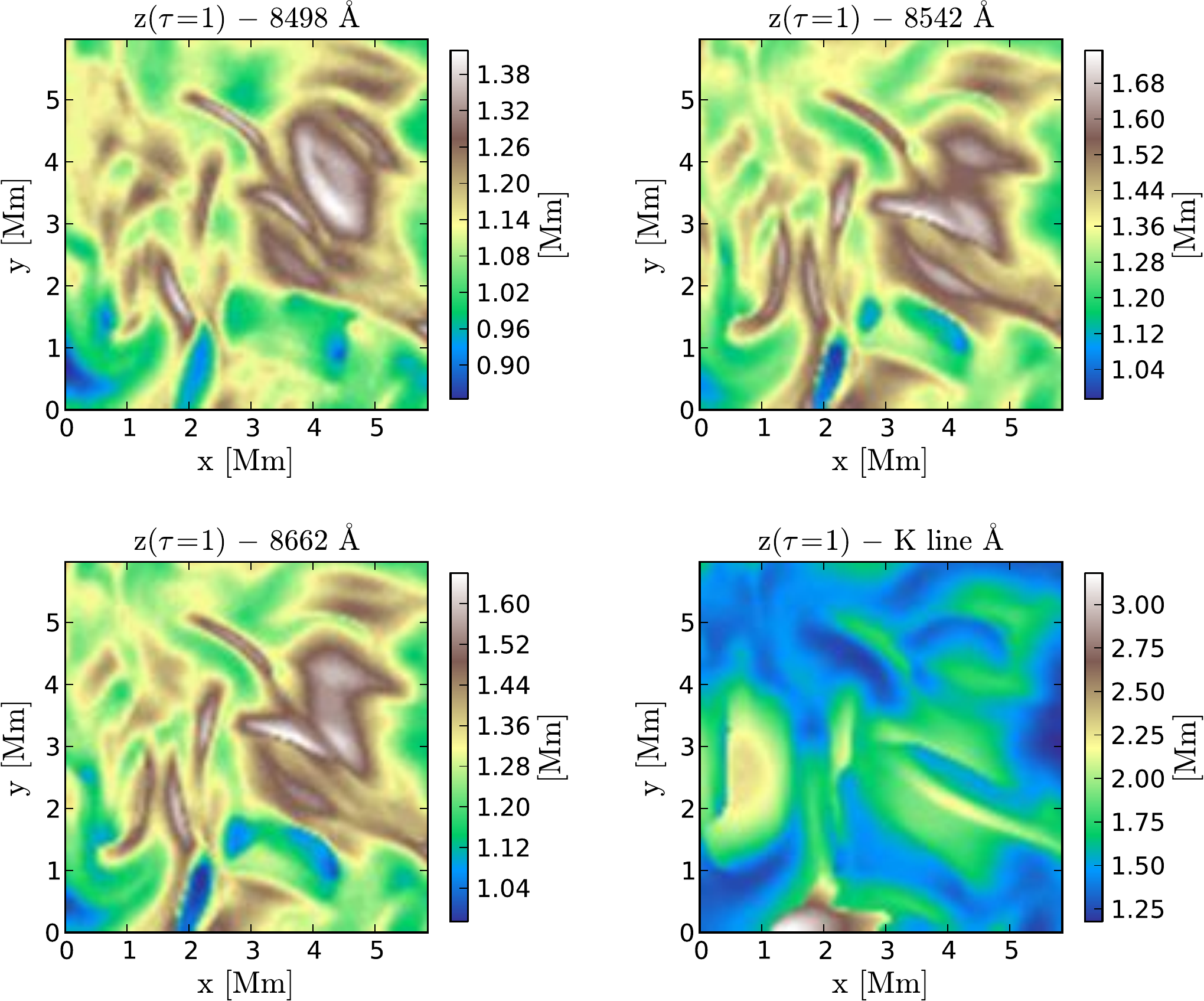}
\caption{\small{Heights of $\tau_{\nu_0}=1$ for the spectral lines under consideration. We can use the K line picture to identify expanded or compressed atmospheres because there is a correlation between this feature and the heights where the K line optical depth is unity.} }
\label{fig:toneh}
\end{figure}
\newpage
\begin{figure}[h!]
        \centering
%        \begin{subfigure}{\textwidth}
                \begin{center}$
                \begin{array}{cc}
                  \includegraphics[width=0.75\textwidth]{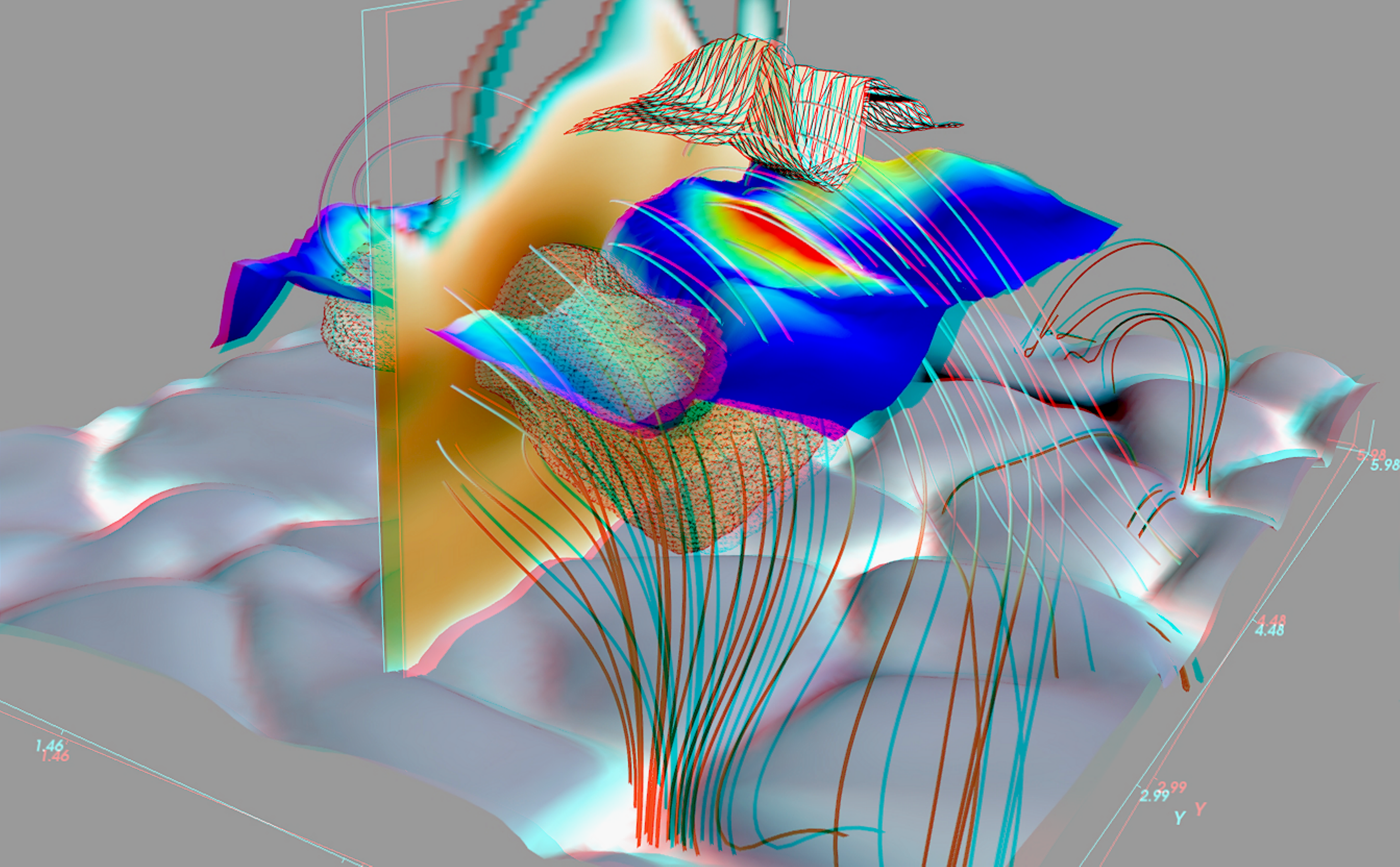} \\
                  \includegraphics[width=0.75\textwidth]{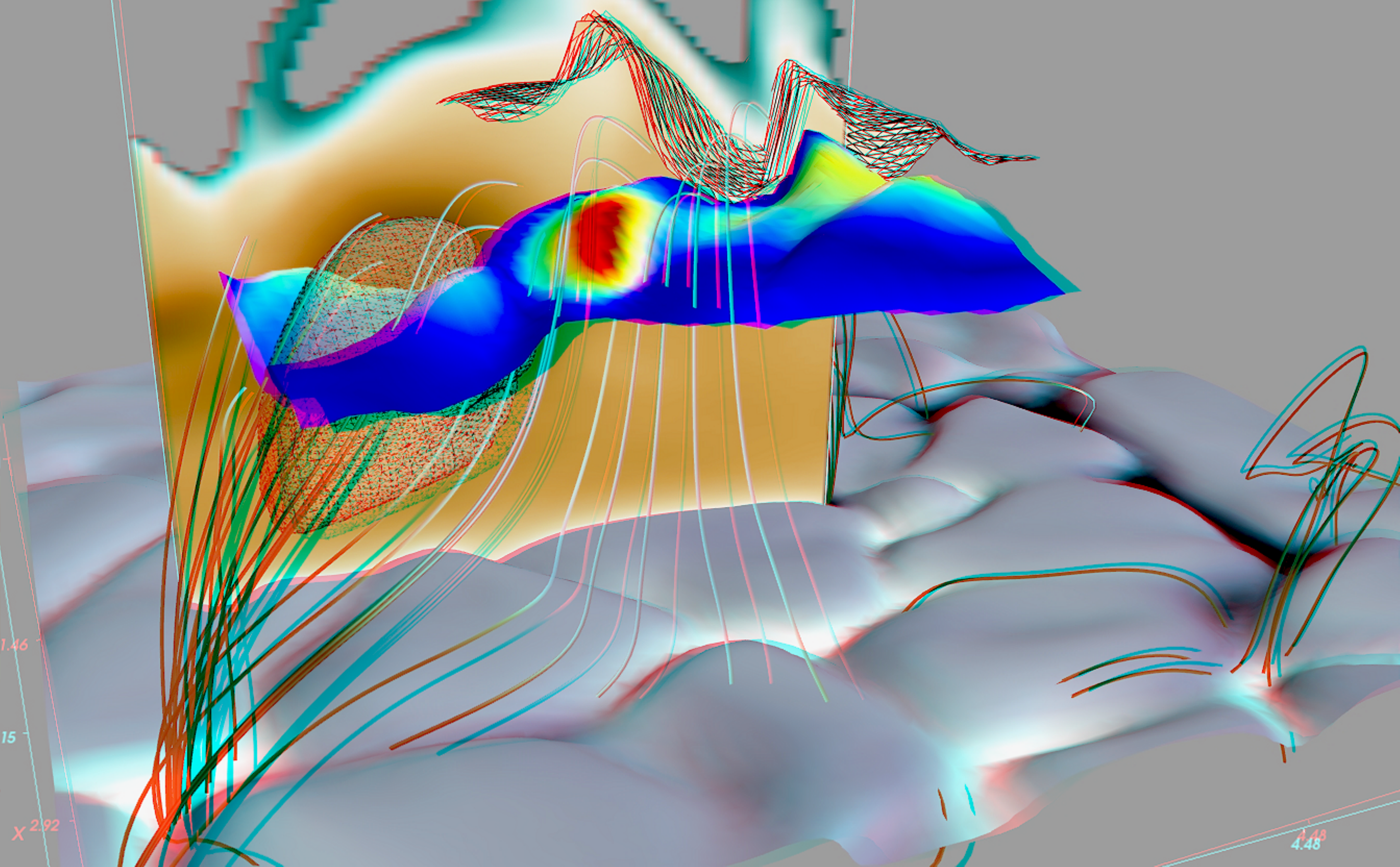} 
                \end{array}$
              \end{center}
 %       \end{subfigure}
        \caption{Similar to Fig. \ref{fig:stereo2} but in a stereographic representation. Cyan-red glasses are
          required to see this figures correctly.
          }\label{fig:stereo}
\end{figure}

%% The following command ends your manuscript. LaTeX will ignore any text
%% /Users/Edgar/ASTROFISICA/SOLAR/WORKING/Escribiendo/mybibdesk_1
%%\bibliographystyle{apj} 
%%\bibliography{mybib}

\begin{thebibliography}{44}
\expandafter\ifx\csname natexlab\endcsname\relax\def\natexlab#1{#1}\fi

\bibitem[{{Anusha} \& {Nagendra}(2011)}]{Anusha:2011ab}
{Anusha}, L.~S., \& {Nagendra}, K.~N. 2011, \apj, 738, 116

\bibitem[{{Anusha} {et~al.}(2011){Anusha}, {Nagendra}, {Bianda}, {Stenflo},
  {Holzreuter}, {Sampoorna}, {Frisch}, {Ramelli}, \& {Smitha}}]{Anusha:2011aa}
{Anusha}, L.~S., {Nagendra}, K.~N., {Bianda}, M., {Stenflo}, J.~O.,
  {Holzreuter}, R., {Sampoorna}, M., {Frisch}, H., {Ramelli}, R., \& {Smitha},
  H.~N. 2011, \apj, 737, 95

\bibitem[{{Asensio Ramos}(2004)}]{Asensio-Ramos:2004}
{Asensio Ramos}, A. 2004, PhD thesis, Universidad de La Laguna, La Laguna

\bibitem[{{Asensio Ramos}(2014)}]{Asensio-Ramos:2014aa}
---. 2014, \aap, 563, A114

\bibitem[{{Bianda} {et~al.}(2011){Bianda}, {Ramelli}, {Anusha}, {Stenflo},
  {Nagendra}, {Holzreuter}, {Sampoorna}, {Frisch}, \&
  {Smitha}}]{michele2011cai}
{Bianda}, M., {Ramelli}, R., {Anusha}, L.~S., {Stenflo}, J.~O., {Nagendra},
  K.~N., {Holzreuter}, R., {Sampoorna}, M., {Frisch}, H., \& {Smitha}, H.~N.
  2011, \aap, 530, L13

\bibitem[{{Bommier}(1997)}]{Bommier:1997ab}
{Bommier}, V. 1997, \aap, 328, 726

\bibitem[{{Carlin} {et~al.}(2013){Carlin}, {Asensio Ramos}, \& {Trujillo
  Bueno}}]{Carlin:2013aa}
{Carlin}, E.~S., {Asensio Ramos}, A., \& {Trujillo Bueno}, J. 2013, \apj, 764,
  40

\bibitem[{{Carlin} {et~al.}(2012){Carlin}, {Manso Sainz}, {Asensio Ramos}, \&
  {Trujillo Bueno}}]{carlin12}
{Carlin}, E.~S., {Manso Sainz}, R., {Asensio Ramos}, A., \& {Trujillo Bueno},
  J. 2012, ApJ, 751, 5

\bibitem[{{Casini} {et~al.}(2014){Casini}, {Landi Degl'Innocenti}, {Manso
  Sainz}, {Landi Degl'Innocenti}, \& {Landolfi}}]{Casini:2014aa}
{Casini}, R., {Landi Degl'Innocenti}, M., {Manso Sainz}, R., {Landi
  Degl'Innocenti}, E., \& {Landolfi}, M. 2014, \apj, 791, 94

\bibitem[{{de la Cruz Rodr{\'{\i}}guez} {et~al.}(2013){de la Cruz
  Rodr{\'{\i}}guez}, {De Pontieu}, {Carlsson}, \& {Rouppe van der
  Voort}}]{de-la-Cruz-Rodriguez:2013aa}
{de la Cruz Rodr{\'{\i}}guez}, J., {De Pontieu}, B., {Carlsson}, M., \& {Rouppe
  van der Voort}, L.~H.~M. 2013, \apjl, 764, L11

\bibitem[{{De la Cruz Rodr\'iguez} {et~al.}(2012){De la Cruz Rodr\'iguez},
  {Socas-Navarro}, {Carlsson}, \& {Leenaarts}}]{delacruz12}
{De la Cruz Rodr\'iguez}, J., {Socas-Navarro}, H., {Carlsson}, M., \&
  {Leenaarts}, J. 2012, Astronomy and Astrophysics, 543, 34

\bibitem[{{Derouich} {et~al.}(2007){Derouich}, {Trujillo Bueno}, \& {Manso
  Sainz}}]{Derouich:2007aa}
{Derouich}, M., {Trujillo Bueno}, J., \& {Manso Sainz}, R. 2007, \aap, 472, 269

\bibitem[{{Hansteen} {et~al.}(2007){Hansteen}, {Carlsson}, \&
  {Gudiksen}}]{Hansteen:2007}
{Hansteen}, V.~H., {Carlsson}, M., \& {Gudiksen}, B. 2007, in Astronomical
  Society of the Pacific Conference Series, Vol. 368, The Physics of
  Chromospheric Plasmas, ed. P.~{Heinzel}, I.~{Dorotovi{\v c}}, \& R.~J.
  {Rutten}, 107

\bibitem[{{Jefferies} {et~al.}(1989){Jefferies}, {Lites}, \&
  {Skumanich}}]{Jefferies:1989aa}
{Jefferies}, J., {Lites}, B.~W., \& {Skumanich}, A. 1989, \apj, 343, 920

\bibitem[{{Judge} {et~al.}(2010){Judge}, {Kn{\"o}lker}, {Schmidt}, \&
  {Steiner}}]{Judge:2010ac}
{Judge}, P., {Kn{\"o}lker}, M., {Schmidt}, W., \& {Steiner}, O. 2010, \apj,
  720, 776

\bibitem[{{Khomenko} \& {Collados}(2012)}]{Khomenko:2012aa}
{Khomenko}, E., \& {Collados}, M. 2012, \apj, 747, 87

\bibitem[{{Landi Degl'Innocenti}(1984)}]{Landi-DeglInnocenti:1984}
{Landi Degl'Innocenti}, E. 1984, Sol. Phys., 91, 1

\bibitem[{{Landi Degl'Innocenti} \& {Landi
  Degl'Innocenti}(1973)}]{Landi-DeglInnocenti:1973}
{Landi Degl'Innocenti}, E., \& {Landi Degl'Innocenti}, M. 1973, Sol. Phys., 31,
  319

\bibitem[{{Landi Degl'Innocenti} \& {Landolfi}(2004)}]{ll04}
{Landi Degl'Innocenti}, E., \& {Landolfi}, M. 2004, Polarization in Spectral
  Lines (Kluwer Academic Publishers)

\bibitem[{{Leenaarts} {et~al.}(2009){Leenaarts}, {Carlsson}, {Hansteen}, \&
  {Rouppe van der Voort}}]{Leenaarts:2009}
{Leenaarts}, J., {Carlsson}, M., {Hansteen}, V., \& {Rouppe van der Voort}, L.
  2009, \apjl, 694, L128

\bibitem[{{Linsky} \& {Avrett}(1970)}]{Linsky:1970aa}
{Linsky}, J.~L., \& {Avrett}, E.~H. 1970, \pasp, 82, 169

\bibitem[{{Manso Sainz} \& {Trujillo Bueno}(2003a)}]{manso_trujillo03a}
{Manso Sainz}, R., \& {Trujillo Bueno}, J. 2003a, Physical Review Letters, 91,
  111102

\bibitem[{{Manso Sainz} \& {Trujillo Bueno}(2003b)}]{Manso-Sainz:2003}
{Manso Sainz}, R., \& {Trujillo Bueno}, J. 2003b, in Astronomical Society of
  the Pacific Conference Series, Vol. 307, Solar Polarization, ed.
  J.~{Trujillo-Bueno} \& J.~{Sanchez Almeida}, 251

\bibitem[{{Manso Sainz} \& {Trujillo Bueno}(2010)}]{manso10}
---. 2010, ApJ, 722, 1416

\bibitem[{{Martinez Pillet} {et~al.}(1990){Martinez Pillet}, {Garcia Lopez},
  {del Toro Iniesta}, {Rebolo}, {Vazquez}, {Beckman}, \&
  {Char}}]{Martinez-Pillet:1990aa}
{Martinez Pillet}, V., {Garcia Lopez}, R.~J., {del Toro Iniesta}, J.~C.,
  {Rebolo}, R., {Vazquez}, M., {Beckman}, J.~E., \& {Char}, S. 1990, \apjl,
  361, L81

\bibitem[{{Mart{\'{\i}}nez-Sykora} {et~al.}(2012){Mart{\'{\i}}nez-Sykora}, {De
  Pontieu}, \& {Hansteen}}]{Martinez-Sykora:2012aa}
{Mart{\'{\i}}nez-Sykora}, J., {De Pontieu}, B., \& {Hansteen}, V. 2012, \apj,
  753, 161

\bibitem[{{Mihalas}(1978)}]{Mihalas:1978}
{Mihalas}, D. 1978, in Stellar Atmospheres, Vol. 455

\bibitem[{{Pietarila} {et~al.}(2007){Pietarila}, {Socas-Navarro}, \&
  {Bogdan}}]{Pietarila:2007aa}
{Pietarila}, A., {Socas-Navarro}, H., \& {Bogdan}, T. 2007, \apj, 670, 885

\bibitem[{{Reardon} {et~al.}(2013){Reardon}, {Tritschler}, \&
  {Katsukawa}}]{Reardon:2013aa}
{Reardon}, K., {Tritschler}, A., \& {Katsukawa}, Y. 2013, \apj, 779, 143

\bibitem[{{Socas-Navarro} {et~al.}(2006){Socas-Navarro}, {Elmore}, {Pietarila},
  {Darnell}, {Lites}, {Tomczyk}, \& {Hegwer}}]{Socas-Navarro:2006}
{Socas-Navarro}, H., {Elmore}, D., {Pietarila}, A., {Darnell}, A., {Lites},
  B.~W., {Tomczyk}, S., \& {Hegwer}, S. 2006, Sol. Phys., 235, 55

\bibitem[{{Stenflo}(1994)}]{Stenflo:1994}
{Stenflo}, J.~O. 1994, Solar Magnetic Fields. Polarized Radiation Diagnostics
  (Dordrecht: Kluwer Academic Publishers)

\bibitem[{{Stenflo}(2003)}]{Stenflo:2003}
{Stenflo}, J.~O. 2003, in Solar Polarization 3, ed. J.~{Trujillo Bueno} \&
  J.~{S\'anchez Almeida}, ASP Conf. Ser. Vol 307, 385

\bibitem[{{Stenflo}(2006)}]{Stenflo:2006aa}
{Stenflo}, J.~O. 2006, in Astronomical Society of the Pacific Conference
  Series, Vol. 358, Astronomical Society of the Pacific Conference Series, ed.
  R.~{Casini} \& B.~W. {Lites}, 215

\bibitem[{{Stenflo} \& {Keller}(1997)}]{Stenflo:1997}
{Stenflo}, J.~O., \& {Keller}, C.~U. 1997, A\&A, 321, 927

\bibitem[{{Trujillo Bueno}(2001)}]{Trujillo-Bueno:2001aa}
{Trujillo Bueno}, J. 2001, in Astronomical Society of the Pacific Conference
  Series, Vol. 236, Advanced Solar Polarimetry -- Theory, Observation, and
  Instrumentation, ed. M.~{Sigwarth}, 161

\bibitem[{{Trujillo Bueno}(2003{\natexlab{a}})}]{Trujillo-Bueno:2003a}
{Trujillo Bueno}, J. 2003{\natexlab{a}}, ASPC, 288, 551

\bibitem[{{Trujillo Bueno}(2003{\natexlab{b}})}]{Trujillo-Bueno:2003aa}
{Trujillo Bueno}, J. 2003{\natexlab{b}}, in Astronomical Society of the Pacific
  Conference Series, Vol. 307, Solar Polarization, ed. J.~{Trujillo-Bueno} \&
  J.~{Sanchez Almeida}, 407

\bibitem[{{Trujillo Bueno} {et~al.}(2002){Trujillo Bueno}, {Landi
  Degl'Innocenti}, {Collados}, {Merenda}, \& {Manso
  Sainz}}]{Trujillo-Bueno:2002a}
{Trujillo Bueno}, J., {Landi Degl'Innocenti}, E., {Collados}, M., {Merenda},
  L., \& {Manso Sainz}, R. 2002, \nat, 415, 403

\bibitem[{{Trujillo Bueno} \& {Manso Sainz}(1999)}]{Trujillo-Bueno:1999aa}
{Trujillo Bueno}, J., \& {Manso Sainz}, R. 1999, apj, 516, 436

\bibitem[{{Trujillo Bueno} {et~al.}(2004){Trujillo Bueno}, {Shchukina}, \&
  {Asensio Ramos}}]{Trujillo-Bueno:2004}
{Trujillo Bueno}, J., {Shchukina}, N., \& {Asensio Ramos}, A. 2004, Nature,
  430, 326

\bibitem[{{Uitenbroek}(1989)}]{Uitenbroek:1989aa}
{Uitenbroek}, H. 1989, \aap, 213, 360

\bibitem[{{Uitenbroek}(2006)}]{Uitenbroek:2006aa}
{Uitenbroek}, H. 2006, in Astronomical Society of the Pacific Conference
  Series, Vol. 354, Solar MHD Theory and Observations: A High Spatial
  Resolution Perspective, ed. J.~{Leibacher}, R.~F. {Stein}, \&
  H.~{Uitenbroek}, 313

\bibitem[{{{\v S}t{\v e}p{\'a}n} \& {Trujillo Bueno}(2013)}]{Stepan:2013aa}
{{\v S}t{\v e}p{\'a}n}, J., \& {Trujillo Bueno}, J. 2013, \aap, 557, A143

\bibitem[{{W{\"o}ger} {et~al.}(2009){W{\"o}ger}, {Wedemeyer-B{\"o}hm},
  {Uitenbroek}, \& {Rimmele}}]{Woger:2009aa}
{W{\"o}ger}, F., {Wedemeyer-B{\"o}hm}, S., {Uitenbroek}, H., \& {Rimmele},
  T.~R. 2009, \apj, 706, 148

\end{thebibliography}

\end{document}